\documentclass[10pt, prb, aps, twocolumn, showpacs, citeautoscript, floatfix, reprint, amsmath, amssymb, notitlepage]{revtex4-1}
\usepackage{booktabs}
\usepackage[utf8]{inputenc}
\usepackage[T1]{fontenc}
\usepackage[english]{babel}
\usepackage{graphicx}
\usepackage{rotating}
\usepackage{mathtools}
\usepackage{amsfonts}
\usepackage{dcolumn} 
\usepackage{bm} 
\usepackage{xcolor}
\usepackage{latexsym}
\usepackage{wasysym}
\usepackage[colorlinks, citecolor=blue, linkcolor=blue, urlcolor=blue]{hyperref}
\usepackage[nomain, acronym, shortcuts]{glossaries}
\usepackage{tikz}
\usepackage{environ}
\usepackage{csquotes}
\glsdisablehyper
\usetikzlibrary{external}
\bibpunct{[}{]}{,}{n}{}{}

\begin{document}

\title{Theory of in-plane current induced spin torque in metal/ferromagnet bilayers}

\newcommand{\affeth}{\affiliation{Institute for Theoretical Physics, ETH Zurich, 8093 Zurich, Switzerland}}

\author{Kohei Sakanashi}\affeth 
\author{Manfred Sigrist}\affeth
\author{Wei Chen}\affeth 

\date{\today}

\begin{abstract}

Using a semiclassical approach that simultaneously incorporates the spin Hall effect (SHE), spin diffusion, quantum well states, and interface spin-orbit coupling (SOC), we address the interplay of these mechanisms as the origin of the in-plane current induced spin torque observed in the normal metal/ferromagnetic metal bilayer thin films. Focusing on the bilayers with a ferromagnet much thinner than its spin diffusion length, such as Pt/Co with $\sim 10$nm thickness, our approach addresses simultaneously the two contributions to the spin torque, namely the spin-transfer torque (SHE-STT) due to SHE induced spin injection, and the spin-orbit torque (SOT) due to SOC induced spin accumulation. The SOC produces an effective magnetic field at the interface, hence it modifies the angular momentum conservation expected for the SHE-STT. The SHE induced spin voltage and the interface spin current are mutually dependent, hence are solved in a self-consistent manner. In addition, the spin transport mediated by the quantum well states may be responsible for the experimentally observed rapid variation of the spin torque with respect to the thickness of the ferromagnet.

\end{abstract}



\maketitle

\section{Introduction}

Ever since the spin-transfer torque (STT) was proposed for the current perpendicular to the plane (CPP) geometry of magnetic heterostructures\cite{Berger96,Slonczewski96}, current induced spin torque has become a major topic in the spintronic research\cite{Stiles02,Zhang02,Stiles06,Ralph08}, as it demonstrates the feasibility of electrical control of magnetization dynamics. To improve the efficiency of magnetization switching, one often aims at reducing the volume or thickness of the magnetic component of the device, which generally enhances the quantum effect on the magnetization dynamics, especially when the thickness of the heterostructures is reduced to the nanometer range. One particularly promising system, owing to its simplicity in manufacturing, is the normal metal/ferromagnetic metal (NM/FMM) bilayer thin films, each layer of thickness of few nanometers, and the NM is a heavy metal such as Pt or Ta\cite{Miron10,Miron11,Liu12,Liu12_2,Garello13}. The intriguing feature discovered in these thin films is that, in contrast to the CPP configuration, a spin torque manifests itself in the current-in-plane (CIP) configuration, whose origin has been attributed to at least the following two mechanisms.

The first is the so-called spin-orbit torque (SOT) originally proposed for a two dimensional system subject to the inversion symmetry breaking in the out-of-plane direction, in which a Rashba spin-orbit coupling (SOC) is anticipated\cite{Obata08,Manchon08,Manchon09,Matos-Abiague09,Haney10,Gambardella11}. An in-plane current induces a spin accumulation in these systems, which then exerts a torque on the magnetization due to the exchange coupling between the conduction electron spin and the magnetization. The interface between NM and FMM obviously breaks the inversion symmetry, hence the SOT is expected to manifest in the CIP configuration. The second mechanism comes from the spin Hall effect (SHE) in the NM\cite{Hirsch99,Murakami03,Sinova04,Sinova15}, in which an in-plane charge current causes a spin current in the transverse direction, i.e., flowing out-of-plane but polarized in-plane. This spin current causes a spin injection into the FMM, resulting in a spin torque. This contribution to the spin torque is frequently called spin Hall effect spin-transfer torque (SHE-STT).

On top of these two mechanisms, the seemingly simple NM/FMM bilayer in reality hosts a number of complexly intertwined features, including anisotropic magnetoresistance (AMR)\cite{McGuire75}, giant magnetoresistance (GMR)-like effects\cite{Avci15,Zhang17}, spin diffusion\cite{Zhang00,Chen13,Ok17}, quantum well states\cite{Moras15,Carbone16}, anomalous Hall effect(AHE)\cite{Pugh53,Sinitsyn08,Nagaosa10}, Berry phase\cite{Sinova04,Kurebayashi14}, anisotropy field, Oersted field, as well as more practical issues such as magnetic domains and spin dependent scattering. A unified theory that can take into account all these effects to explain the observed spin torque has yet been formulated, and may be too complicated to analyze. Thus attempts to formulate a theoretical description for the spin torque have been focusing on some of these mechanisms that are thought to be most relevant. Using different approaches, a damping-like spin torque is attributed to the combined effect of interface SOC and spin relaxation\cite{Kim12,Wang12} or spin dependent scattering\cite{Pesin12}. A semiclassical theory that combines the drift-diffusion approach with the Boltzmann equation has also been proposed\cite{Haney13_2}, from which the spatial profile of spin voltage under the influence of spin diffusion, SHE, and interface Rashba SOC are obtained. A first principle calculation concerning SOT and realistic band structures, without incorporating SHE, has also been performed\cite{Haney13}. The dependence of the spin torque on the magnetization direction\cite{Garello13,Qiu14} has been attributed to the anisotropic spin relaxation rates\cite{Pauyac13} and the Berry phase effect\cite{Lee15}.


Here we present a semiclassical theory for the NM/FMM bilayer in the limit that the FMM is much thinner than its spin diffusion length. We consider this limit to be relevant to most experiments that use ferromagnetic materials whose bulk spin diffusion length far exceeds the film thickness, such as Co and Ni, but not applicable to materials with short spin diffusion length such as permalloy\cite{Bass07}. Our treatment incorporates four of the aforementioned intertwined mechanisms and the appropriate theoretical methods to describe them, namely the spin diffusion equation for the SHE and spin diffusion in the NM\cite{Zhang00,Chen13,Ok17}, the quantum Boltzmann approach for the interface Rashba SOC\cite{Manchon08,Gambardella11}, and the quantum tunneling theory for the SHE induced spin injection mediated by the quantum well state\cite{Chen15_STT,Chen16_quantum_tunneling,Ok17}. Our goal is not to quantify the spin torque for a specific set of bilayers, since we made several approximations to simplify our calculations, including the parabolic band and the sharp Rashba interface approximation. Rather, we aim at extracting quantitative statements for generic bilayers considering how each material parameter influences the SHE-STT part and the SOT part of the spin torque, which may help to engineer the spin torque for practical applications. We particularly focus on four of the system parameters, namely the FMM thickness $l_{FM}$, NM thickness $l_{N}$, exchange coupling $J_{sd}$, and interface SOC $\alpha_{R}k_{F}$, over which the experimentalist may have control in reality, and show how their interplay may explain a number of experimental observations, such as the rapid variation of the spin torque with respect to the FMM thickness\cite{Kim13}. Although we limit our discussion to a single domain FMM, the two parts of the spin torque are expected to contribute to the current induced domain wall motion observed in the NM/FMM bilayers where the FMM consists of multiple domains\cite{Lee11,Miron11_2,Emori13}, hence we also draw relevance to several phenomena therein, especially the dependence on the layer thickness.

The article is structured in the following manner. In Sec.~\ref{sec:NMFMM_theory}, we formulate a quantum tunneling theory for the interface spin current, which is then combined with the spin diffusion equation to self-consistently solve for the interface spin voltage and the interface spin current. The SHE-STT and the issue of angular momentum conservation are then addressed. The SOT is then calculated from the interface SOC, which joins the SHE-STT to give the total spin torque. Numerical results using realistic parameters are then presented. Sec.~\ref{sec:Conclusions} gives a summary of the features revealed by our approach.

\begin{figure}
\centering
\includegraphics[clip=true,width=0.8\columnwidth]{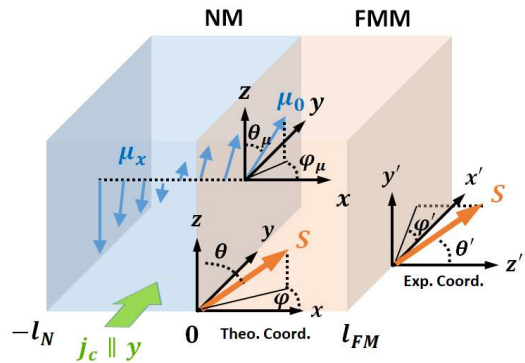}
\caption{ (Color online) Schematics of the NM/FMM bilayer. The quantum tunneling theory adopts the coordinate $(x,y,z)$ where out-of-plane direction is ${\hat {\bf x}}$, in contrast to the experimental coordinate $(x^{\prime},y^{\prime},z^{\prime})$ where out-of-plane direction is denoted by ${\hat {\bf z}^{\prime}}$. The FMM magnetization expressed in these coordinates is ${\bf S}=(S^{x},S^{y},S^{z})=S(\sin\theta\cos\varphi,\sin\theta\sin\varphi,\cos\theta)$ and ${\bf S}=(S^{x\prime},S^{y\prime},S^{z\prime})=S(\sin\theta^{\prime}\cos\varphi^{\prime},\sin\theta^{\prime}\sin\varphi^{\prime},\cos\theta^{\prime})$. Blue arrows indicate the spatial profile of SHE induced spin voltage ${\boldsymbol\mu}_{x}$ in the NM under the influence of spin diffusion and interface spin current ${\boldsymbol j}_{0-}$. The spin voltage at the interface is denoted by ${\boldsymbol\mu}_{0}=(\mu_{0}^{x},\mu_{0}^{y},\mu_{0}^{z})=|{\boldsymbol\mu}_{0}|(\sin\theta_{\mu}\cos\varphi_{\mu},\sin\theta_{\mu}\sin\varphi_{\mu},\cos\theta_{\mu})$. The thickness of NM and FMM are labeled by $l_{N}$ and $l_{FM}$, respectively, and the interfac is located at $x=0$. } 
\label{fig:NMFMM_schematics}
\end{figure}

\section{Theoretical description of NM/FMM bilayer \label{sec:NMFMM_theory}}

\subsection{Self-consistent treatment of spin voltage and SHE-STT \label{sec:GrGi_theory}}

We first present a semiclassical formalism that self-consistently solves for the spin voltage ${\boldsymbol\mu}_{x}$, spin current ${\boldsymbol j}_{x}$, and SHE-STT ${\boldsymbol\tau}_{STT}$. Our treatment is based on the assumption that the spin dynamics in the FMM is purely quantum since it is much thinner than its spin diffusion length $\lambda\gg l_{FM}$, and the spin dynamics in the NM is purely diffusive except near the interface where the spin voltage causes spin injection. We define the out-of-plane direction to be ${\hat{\bf x}}$ and the direction of charge current to be ${\hat{\bf y}}$\cite{Berger96,Chen15_STT,Chen16_quantum_tunneling,Ok17}, as shown in Fig.~\ref{fig:NMFMM_schematics}, in contrast to the experimental convention of coordinates which are denoted by $(x^{\prime},y^{\prime},z^{\prime})$. Owing to the translational invariance in the $yz$-plane, the spin voltage and the spin current are only functions of out-of-plane coordinate $x$. To incorporate the translational invariance, we perform a separation of variables for the Hamiltonian and the wave function that describe the entire FMM and the NM near the interface
\begin{eqnarray}
H_{N/F}&=&H_{N/F,x}+H_{N/F,yz}\;,
\nonumber \\
\Psi_{N/F}&=&\psi_{N/F}(x)\psi_{N/F}(y,z)\;.
\end{eqnarray}
Within the parabolic band approximation, the Hamiltonian in the ${\hat{\bf x}}$ direction is
\begin{eqnarray}
H_{N,x} &=& \frac{\hbar^{2}k_{x}^{2}}{2m} - \frac{{\boldsymbol\mu}_{0}\cdot{\boldsymbol\sigma}}{2} \;\;\; (x \apprle 0) \: ,
\nonumber \\
H_{F,x} &=& \frac{\hbar^{2}k_{x}^{2}}{2m} +J_{sd}\;{\bf S}\cdot{\boldsymbol\sigma} \;\;\; (0 \leq x \leq l_{FM})\;.
\nonumber \\
H_{I,x} &=& \alpha_R({\bf k}\times \hat{{\bf x}}) \cdot {\boldsymbol \sigma}\delta(x)a\;, 
\end{eqnarray}
where we approximate the spin voltage in the vincinity of the interface by its interface value ${\boldsymbol\mu}_{x\apprle 0}\approx{\boldsymbol\mu}_{0}=|{\boldsymbol\mu}_{0}|(\cos \varphi_{\mu} \sin \theta_{\mu},\sin \varphi_{\mu} \sin \theta_{\mu},\cos \theta_{\mu})$. The spin voltage splits the degeneracy of spin $\sigma=\left\{\uparrow,\downarrow\right\}$ quantized along ${\boldsymbol\mu}_{0}$. The $J_{sd}$ is the exchange coupling between conduction electron spin and the magnetization ${\boldsymbol S}=S(\cos \varphi \sin \theta,\sin \varphi \sin \theta,\cos \theta)$, which is set to be negative $J_{sd}<0$ such that the conduction electron spin and the magnetization tend to be parallel. The $\alpha_{R}$ term is the Rashba SOC assumed to be sharply confined at the interface\cite{Haney13}, and the $\delta$-function that satisfies $\int dx\delta(x)=1$ is multiplied by Fermi wave length $a$ to keep track of the dimension. The corresponding wave functions in the out-of-plane direction are 
\begin{eqnarray}
   \psi_{N}(x) &=& 
   (Ae^{ik_{0 \uparrow}x} + Be^{-ik_{0 \uparrow}x}) \left( \begin{array}{c} e^{-i \varphi_{\mu} /2} \cos \frac{\theta_{\mu}}{2} \\ e^{i \varphi_{\mu} /2} \sin \frac{\theta_{\mu}}{2}  \end{array} \right)
\nonumber \\   
&+&  Ce^{-ik_{0 \downarrow}x} \left( \begin{array}{c} -e^{-i \varphi_{\mu} /2} \sin \frac{\theta_{\mu}}{2} \\ e^{i \varphi_{\mu} /2} \cos \frac{\theta_{\mu}}{2} \end{array} \right)\;, 
\nonumber \\
   \psi_{F}(x) &=& 
   e^{ik_{+}l_{FM}}2i\sin \left[k_{+}(x-l_{FM})\right]D \left( \begin{array}{c} e^{-i \varphi /2} \cos \frac{\theta}{2} \\ e^{i \varphi /2} \sin \frac{\theta}{2} \end{array} \right)
   \nonumber \\ 
    &+& e^{ik_{-}l_{FM}}2i\sin \left[k_{-}(x-l_{FM})\right]E \left( \begin{array}{c} -e^{-i \varphi /2} \sin \frac{\theta}{2} \\ e^{i \varphi /2} \cos \frac{\theta}{2} \end{array} \right),
\nonumber \\
k_{0\sigma}&=&\sqrt{2m(\epsilon_{F}\pm|{\boldsymbol\mu}_{0}|/2)}/\hbar\;,
\nonumber \\
k_{\pm} &=& \sqrt{2m(\epsilon_{F} \mp J_{sd} S)}/ \hbar\;,
\label{NM_FMM_wave_fn}
\end{eqnarray}
where the spinor of $\psi_{N}$ and $\psi_{F}$ are quantized along the interface spin voltage and the magnetization, respectively. Recent angle-resolved photoemission spectroscopy (ARPES) experiments in NM/FMM bilayers unambiguously demonstrate the existence of the exchange-split quantum well states\cite{Moras15,Carbone16}, which is described by the oscillatory wave function $\psi_{F}$ in our formalism. The wave function outside of the FMM $x>l_{FM}$, which is usually an oxide insulator or vacuum, is assumed to vanish for simplicity\cite{Ok17}.

The spin voltage is caused by the nonequilibrium electrons that also have momentum $k_{y}{\hat{\bf y}}\approx k_{F}{\hat{\bf y}}$, since the charge current is flowing along ${\hat{\bf y}}$ direction. Thus the Rashba term reads $H_{I,x}=-\alpha_{R}k_{F}\sigma^{z}\delta(x)a$. Defining $\beta_{R}=2m\alpha_{R}k_{F}a/\hbar^{2}$, the matching conditions at the interface become
\begin{eqnarray}
&&\psi_{F}(0)-\psi_{N}(0)=0\;,
\nonumber \\
&&\partial_{x}\psi_{F}(x)|_{x=0}-\partial_{x}\psi_{N}(x)|_{x=0}=\beta_{R}\sigma^{z}\psi_{N/F}(0)\;,
\label{wave_fn_boundary_condition}
\end{eqnarray}
which are used to solve the coefficients $B\sim E$ in terms of the incident amplitude $A$. The incoming flux is then identified with the interface spin voltage $|A|^{2}=N_{F}|{\boldsymbol\mu}_{0}|/a^{3}$, where $N_{F}$ is the density of states per $a^{3}$. This identification bridges the quantum tunneling formalism above and the spin diffusion equation below.

The interface SOC renders an interesting consequence for the angular momentum conservation\cite{Slonczewski96}. The spin current right before $(x=0-)$ and right after $(x=0+)$ the interface is, using Eq.~(\ref{wave_fn_boundary_condition}), 
\begin{eqnarray}
   \boldsymbol{j}_{0-} &=& \frac{\hbar}{4im} \big[ \psi_{N}^{*} \boldsymbol{\sigma} (\partial_{x} \psi_{N})|_{x=0} - (\partial_{x} \psi_{N}^{*}) \boldsymbol{\sigma} \psi_{N}|_{x=0} \big]
\nonumber \\
&=&\frac{\hbar}{4im} \left[ \psi_{F}^{*} \boldsymbol{\sigma} (\partial_{x} \psi_{F})|_{x=0} - (\partial_{x} \psi_{F}^{*}) \boldsymbol{\sigma} \psi_{F}|_{x=0}\right. 
\nonumber \\
&&\left.-\beta_{R}\psi_{N/F}^{*}\left({\boldsymbol\sigma}\sigma^{z}-\sigma^{z}{\boldsymbol\sigma}\right)\psi_{N/F}\right]
\nonumber \\
&=&{\boldsymbol j}_{0+}+\delta{\boldsymbol j}_{0}\;.
\label{j0p_j0m}
\end{eqnarray}
That is, in the presence of interface SOC, the spin current is not conserved across the interface. Evidently, this is because the SOC effectively acts like a magnetic field at $x=0$, hence changing the polarization of the injected spin current. On the other hand, one may calculate the spin accumulation at position $x$ inside the FMM by $\langle{\boldsymbol\sigma}_{x}\rangle=\psi_{F}^{\ast}(x){\boldsymbol\sigma}\psi_{F}(x)$, and then multiply by cross section unit and integrate along out-of-plane direction to get the total spin accumulation $\overline{\langle{\boldsymbol\sigma}\rangle}=a^{2}\int_{0+}^{l_{FM}}dx\langle{\boldsymbol\sigma}_{x}\rangle$. The SHE-STT is then obtained via Landau-Lifshitz dynamics, and we find that
\begin{eqnarray}
{\boldsymbol\tau}_{STT}=\frac{J_{sd}}{\hbar}\overline{\langle{\boldsymbol\sigma}\rangle}\times{\bf S}=a^{2}{\boldsymbol j}_{0+}\;,
\label{angular_momentum_sonservation}
\end{eqnarray}
meaning that the SHE-STT is equal to the spin current right after the interface ${\boldsymbol j}_{0+}$ but not before ${\boldsymbol j}_{0-}$. This is a very peculiar feature of the CIP configuration with SHE induced out-of-plane spin injection, in contrast to the CPP configuration where angular momentum conservation is satisfied exactly at the interface\cite{Slonczewski96}. This feature is not included in our previous treatments\cite{Chen15_STT,Chen16_quantum_tunneling,Ok17}. Comparing Eqs.~(\ref{j0p_j0m}) and (\ref{angular_momentum_sonservation}), we see that interface SOC changes angular momentum transferred from the NM to FMM, thus engineering interface SOC may also influence the SHE-STT.

We proceed to review the spin diffusion approach that describes the spin voltage in the NM\cite{Zhang00,Kato04}, and how the interface spin current modifies the landscape of the spin voltage\cite{Chen13}. The quantum tunneling theory in Eqs.~(\ref{NM_FMM_wave_fn}) to (\ref{angular_momentum_sonservation}) will be incorporated into this diffusive formalism later. The spin diffusion approach is based on the following properties of the NM: (1) The spin current in NM consists of both the spatial gradient of spin voltage and the bare SHE spin current $j_{SH}=\theta_{SH}\sigma_{c}E/2e$, i.e.,
\begin{eqnarray}
\boldsymbol{j}_{x} &=& - \frac{\sigma_{c}}{4e^{2}} \partial_{x} \boldsymbol{\mu}_{x} + j_{SH} \hat{\boldsymbol{z}} \: ,
\label{jx_mu_E}
\end{eqnarray}
where $\theta_{\textrm{SH}}$ is spin Hall angle, $\sigma_{c}$ is the conductivity of NM, $E=j_{c}/\sigma_{c}$ is the electric field in $y$ direction, and $-e$ is electron charge.
 (2) The spin voltage obeys the spin diffusion equation $\nabla^{2} \boldsymbol{\mu}_{x} = \boldsymbol{\mu}_{x} / \lambda^{2}$, where $\lambda$ is the spin diffusion length, and hence have a general solution
\begin{eqnarray}
{\boldsymbol\mu}_{x}={\bf A}e^{x/\lambda}+{\bf B}e^{-x/\lambda}\;.
\label{spin_voltage_general_solution}
\end{eqnarray}
(3) Spin current vanishes at the edge of NM, ${\boldsymbol j}_{-l_{N}}=0$, which serves as one boundary condition. (4) The spin current at the interface is described by the ${\boldsymbol j}_{0-}$ in Eq.~(\ref{j0p_j0m}), i.e., the spin current close to the interface on the NM side, which serves as another boundary condition. The self-consistent solution satisfying (1)$\sim$(4) gives the landscape of the spin voltage and the spin current\cite{Chen13} 
\begin{eqnarray}
{\boldsymbol\mu}_{x}&=&\tilde{\mu}_{0}\frac{\sinh\left(\frac{2x+l_{N}}{2\lambda}\right)}{\sinh\left(\frac{l_{N}}{2\lambda}\right)}{\hat{\bf z}}-\frac{4e^{2}\lambda}{\sigma_{c}}\frac{\cosh\left(\frac{x+l_{N}}{\lambda}\right)}{\sinh\left(\frac{l_{N}}{\lambda}\right)}{\boldsymbol j}_{0-}\;,
\nonumber \\
{\boldsymbol j}_{x}&=&\left[-\frac{\sigma_{c}\tilde{\mu}_{0}}{4e^{2}\lambda}\frac{\cosh\left(\frac{2x+l_{N}}{2\lambda}\right)}{\sinh\left(\frac{l_{N}}{2\lambda}\right)}
+j_{SH}\right]{\hat{\bf z}}
+\frac{\sinh\left(\frac{x+l_{N}}{\lambda}\right)}{\sinh\left(\frac{l_{N}}{\lambda}\right)}{\boldsymbol j}_{0-}\;,
\nonumber \\
\label{spin_diffusion_equation}
\end{eqnarray}
where $\tilde{\mu}_{0}=2e\lambda\theta_{SH}E\tanh\left(l_{N}/2\lambda\right)$ is the surface spin voltage when the NM is not attached to the FMM. Equations (\ref{jx_mu_E}) $\sim$ (\ref{spin_diffusion_equation}) remain valid even in the presence of SOC, as they only rely on the assumptions (1) $\sim$ (4) which remain true even in the presence of SOC.

Exactly at the interface, the spin voltage satisfies
\begin{eqnarray}
{\boldsymbol\mu}_{0}&=&\tilde{\mu}_{0}{\hat{\bf z}}-\frac{4e^{2}\lambda}{\sigma_{c}}\coth\left(\frac{l_{N}}{\lambda}\right){\boldsymbol j}_{0-}\;,
\end{eqnarray}
so the interface spin voltage is determined by the interface spin current. 
On the other hand, from Eq.~(\ref{NM_FMM_wave_fn}) we see that the interface spin voltage influences the incident momentum $k_{0\sigma}$ and the incident flux $|A|^{2}=N_{F}|{\boldsymbol\mu}_{0}|/a^{3}$, and hence determines the interface spin current ${\boldsymbol j}_{0-}$ in Eq.~(\ref{j0p_j0m}). Thus, they are mutually dependent
\begin{eqnarray}
{\boldsymbol\mu}_{0}={\boldsymbol\mu}_{0}({\boldsymbol j}_{0-})\;,
\nonumber \\
{\boldsymbol j}_{0-}={\boldsymbol j}_{0-}({\boldsymbol\mu}_{0})\;,
\label{mu0_j0_self_consistent}
\end{eqnarray}
which are to be solved self-consistently by iteration at a given set of parameters. After the solution is obtained, which typically converges within few hundred iterations, we put the resulting ${\boldsymbol j}_{0-}$ into Eq.~(\ref{spin_diffusion_equation}) to obtain the spatial profile of ${\boldsymbol\mu}_{x}$ and ${\boldsymbol j}_{x}$ in the NM, and use Eq.~(\ref{j0p_j0m}) to obtain ${\boldsymbol j}_{0+}$ which then gives the SHE-STT according to Eq.~(\ref{angular_momentum_sonservation}). It is convenient to introduce a frequency scale 
\begin{eqnarray}
\omega_{0}=\frac{2e\lambda E\theta_{SH}}{\hbar}\sim 10\;{\rm GHz}\;,
\;\;\;\frac{\hbar\omega_{0}}{\mu_{B}}\sim 100{\rm mT}\;,
\label{omega_0}
\end{eqnarray}
according to the typical values in Table \ref{tab:notations}, and express the STT accordingly
\begin{eqnarray}
{\boldsymbol\tau}_{STT}&=&\omega_{0}\left(\frac{|{\boldsymbol\mu}_{0}|}{2e\lambda E\theta_{SH}}\right)
\nonumber \\
&&\times\left\{\frac{a}{2i|A|^{2}}\left[\psi_{F}^{*} \boldsymbol{\sigma} (\partial_{x} \psi_{F}) - (\partial_{x} \psi_{F}^{*}) \boldsymbol{\sigma} \psi_{F}\right]_{x=0}\right\}\;.
\nonumber \\
\end{eqnarray}
The expression inside the bracket is usually of the order of unity. Thus the STT alone is in the range of GHz, consistent with that measured experimentally.

\begin{table}[ht]
  \centering
  \caption{Summary of the notations and their order of magnitude values when the external charge current is fixed at $j_{c}\sim 10^{11}$A/m$^{2}$.}
  \label{tab:notations}
\begin{ruledtabular}
  \begin{tabular}{ll}
Fermi momentum & $k_{F}\sim 1/a\sim$nm$^{-1}$ \\
Fermin energy & $\epsilon_{F}\sim$ eV \\
Conductivity & $\sigma_{c}\sim 10^{7}$S/m \\
Spin Hall angle & $\theta_{SH}\sim 0.1$ \\
NM spin diffusion length & $\lambda\sim 10$ nm \\
External charge current & $j_{c}\sim 10^{11}$A/m$^{2}$ \\
External field & $E\sim 10^{4}$mkg/Cs$^{2}$ \\
Relaxation time & $\tau\sim 10^{-14}$s \\
Dimensionless magnetization & $S\sim 1$ \\
Bare spin Hall spin current & $j_{SH}\sim 10^{29}/$m$^{2}$s \\
Temperature in SOT integral & $k_{B}T\sim 0.02\;\epsilon_{F}$
  \end{tabular}
\end{ruledtabular}
\end{table}

\subsection{SOC induced spin accumulation and SOT \label{sec:spin_accumulation}}

We proceed to address the other component of the spin torque, namely the SOT\cite{Manchon08,Manchon09,Haney10,Pesin12,Haney13,Gambardella11}. Under the assumption that the parity-breaking potential at the interface is extremely sharp\cite{Haney13}, we isolate the $yz$ plane at $x=0$ and consider the Hamiltonian
\begin{eqnarray}\label{eq:eq100}
H_{I,yz}&=&\delta(x)a\left\{\frac{\hbar^{2}(k_{y}^{2}+k_{z}^{2})}{2m}\right.
\nonumber \\
&&\left.+\left[\alpha_R({\bf k}\times \hat{{\bf x}})+ J_{sd}{\boldsymbol S}
-\frac{{\boldsymbol\mu}_{0}}{2}\right]\cdot{\boldsymbol\sigma}\right\}\;,
\label{interface_Hamiltonian}
\end{eqnarray}
where ${\bf k}=(0,k^{y},k^{z})=k(0,\cos \xi,\sin \xi)$ is the in-plane momentum. Equation (\ref{interface_Hamiltonian}) describes an interface simultaneously under the influence of interface Rashba SOC, exchange coupling, and the interface spin voltage since it is a parity breaking interface proximity to both the NM and the FMM. However, using the typical values for the parameters in Table \ref{tab:notations}, the interface spin voltage is typically $|{\boldsymbol\mu}_{0}|\sim 10^{-5}$eV, which is few orders of magnitude smaller than the SOC $\alpha_{R}k_{F}\sim 0.01$eV and the exchange coupling $J_{sd}\sim 0.1$eV, meaning that interface spin voltage is not crucial for the SOT. In the calculation below we ignore the ${\boldsymbol\mu}_{0}$ term in Eq.~(\ref{interface_Hamiltonian}) for simplicity. This also means that SOT is not influenced by the self-consistent approach to SHE-STT in Sec.~\ref{sec:GrGi_theory}.

Our aim is to calculate the spin accumulation induced by an in-plane current flowing along ${\hat{\bf y}}$ direction using quantum Boltzmann equation. Assuming the relaxation time $\tau$ is the same for the two bands of the Hamiltonian, denoting $f_{{\bf k},\pm}^{0}$ as the equilibrium Fermi distribution for the two bands, and ${\bf E}=E{\hat{\bf y}}$ as the electric field associated with the charge current, the leading order distribution function is
\begin{equation}
g_{{\bf k},\pm} = -\frac{e}{\hbar}{\bf E}\cdot {\bf v}_{k,\pm}\tau\frac{\partial f_{{\bf k},\pm}^0}{\partial E_{{\bf k},\pm}}.
\label{distribution_function}
\end{equation}
from which one calculates the interface spin accumulation via the Boltzmann equation
\begin{eqnarray}
&&\langle\delta{\boldsymbol \sigma}\rangle_{\pm}= \int \frac{d^2{\bf k}}{(2\pi/a)^2} \langle{\boldsymbol \sigma}\rangle_{\pm}g_{k,\pm}\;.
\label{Boltzmann_eq}
\end{eqnarray}
where we denote the spin expectation value of the eigenstate as $\langle{\boldsymbol\sigma}\rangle_{\pm}$. In Appendix \ref{append:Boltzmann_eq}, we provide a formalism that is convenient for numerically calculating $\langle\delta{\boldsymbol \sigma}\rangle_{\pm}$ at any values of $\left\{\alpha_{R}k_{F},J_{sd}\right\}$.
The total spin accumulation is the sum of the contribution from each band $\langle\delta{\boldsymbol\sigma}\rangle=\langle\delta{\boldsymbol\sigma}\rangle_{+}+\langle\delta{\boldsymbol\sigma}\rangle_{-}$, which is multiplied by lattice unit area and $\delta(x)a$ due to Eq.~(\ref{interface_Hamiltonian}) and then integrated along out-of-plane direction to get the total spin accumulation in the whole FMM film $\overline{\langle\delta{\boldsymbol\sigma}\rangle}=a^{2}\int_{0-}^{l_{FM}}dx\langle\delta{\boldsymbol\sigma}\rangle\delta(x)a
=a^{3}\langle\delta{\boldsymbol\sigma}\rangle$.
The SOT then follows from the Landau-Lifshitz dynamics
\begin{eqnarray}
{\boldsymbol\tau}_{SOT}=\frac{J_{sd}}{\hbar}\overline{\langle\delta{\boldsymbol\sigma}\rangle}\times{\bf S}\;.
\label{SOT_expression}
\end{eqnarray}
Since ${\boldsymbol\tau}_{SOT}$ comes from the interface SOC, it does not strongly depend on the FMM thickness. From Eq.~(\ref{deltasigma_integral}), we see that it is convenient to introduce a frequency scale 
\begin{eqnarray}
\omega_{1}=\frac{\epsilon_{F}^{3/2}e\tau E\sqrt{2m}}{4\pi^{2}\hbar^{3}a}\sim 10\;{\rm GHz}\;,
\;\;\;\frac{\hbar\omega_{1}}{\mu_{B}}\sim 100{\rm mT}\;,\;\;
\label{omega_1}
\end{eqnarray}
after using the typical values in Table \ref{tab:notations}, and express SOT accordingly 
\begin{eqnarray}
{\boldsymbol\tau}_{SOT}=\left(\frac{J_{sd}}{\epsilon_{F}}\right)\omega_{1}{\bf I}\times{\bf S}\;,
\end{eqnarray}
where ${\bf I}$ is the dimensionless integral part of Eq.~(\ref{deltasigma_integral}) whose numerical value is typically $\sim{\cal O}(1)$. Comparing Eqs.~(\ref{omega_0}) and (\ref{omega_1}), one sees that STT and SOT have comparable magnitudes, and both are in the GHz regime, in accordance with that measured experimentally.

\subsection{Numerical results\label{sec:numerical_results}}

Following the experimental convention, we define the field-like ${\hat{\bf S}}\times{\hat{\bf z}}$ and damping-like ${\hat{\bf S}}\times({\hat{\bf S}}\times{\hat{\bf z}})$ direction from ${\hat{\bf z}}$ and ${\hat{\bf S}}$. Note that they are different from that defined from ${\boldsymbol\mu}_{0}$ and ${\hat{\bf S}}$\cite{Chen15_STT,Chen16_quantum_tunneling,Ok17}. The total spin torque, i.e., the sum of the SHE-STT discussed in Sec.~\ref{sec:GrGi_theory} and the SOT discussed in Sec.~\ref{sec:spin_accumulation}, is then projected into these two components 
\begin{eqnarray}
{\boldsymbol\tau}&=&\frac{d{\bf S}}{dt}={\boldsymbol\tau}_{STT}+{\boldsymbol\tau}_{SOT}
\nonumber \\
&=&\tau_{f}{\hat{\bf S}}\times{\hat{\bf z}}+\tau_{d}{\hat{\bf S}}\times({\hat{\bf S}}\times{\hat{\bf z}})\;.
\end{eqnarray}
Below we discuss the numerical result of ${\boldsymbol\mu}_{x}$ and ${\boldsymbol j}_{x}$, as well as the dependence of $\tau_{f}$ and $\tau_{d}$ on various material parameters.

\begin{figure}
\centering
\includegraphics[clip=true,width=0.99\columnwidth]{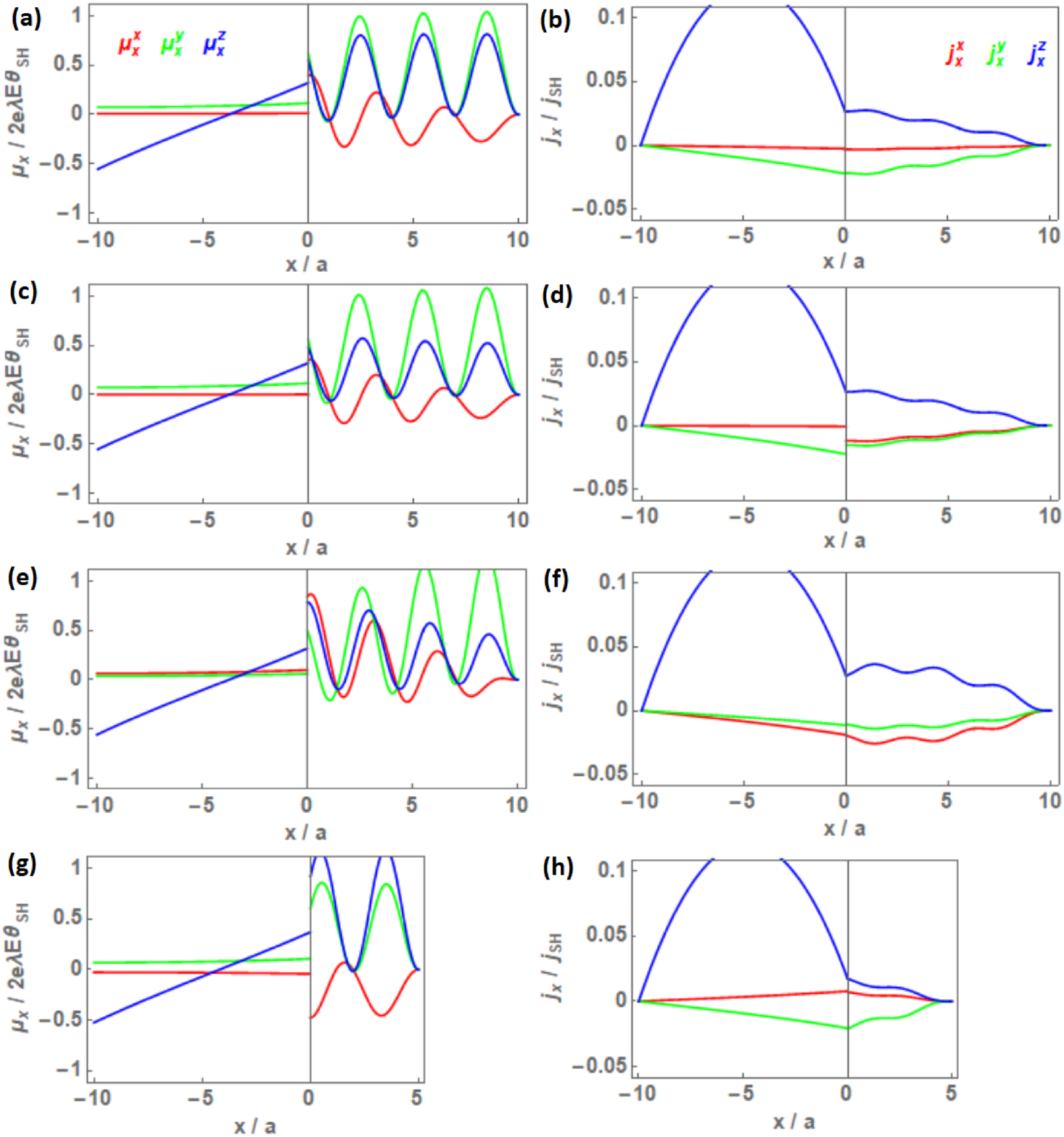}
\caption{ (Color online) Spin voltage (left column, in units of $2e\lambda E\theta_{SH}\sim 10^{-5}$eV) and spin current (right column, in units of $j_{SH}\sim 10^{29}$/sm$^{2}$) in the NM/FMM bilayer. Each color represnts a spin polarization. (a) and (b) are at $l_{FM}=l_{N}=10a$, $J_{sd}/\epsilon_{F}=-0.1$, $\alpha_{R}k_{F}/\epsilon_{F}=0.01$, $\theta=0.3\pi$, and $\varphi=0.3\pi$. These parameters serve as the reference point for the following plots. In (c) and (d) the interface SOC is increased to $\alpha_{R}k_{F}/\epsilon_{F}=0.2$, causing an obvious discontinuity of spin current at the interface as described by Eq.~(\ref{j0p_j0m}). In (e) and (f) the exchange coupling is increased to $J_{sd}/\epsilon_{F}=-0.2$, changing the pattern of spin voltage in the FMM due to quantum interference effect of the spin transport. In (g) and (h) the FMM thickness is reduced to $l_{FM}=5a$, showing that the spin voltage inside FMM is dramatically influenced by $l_{FM}$, again due to the quantum interference effect. } 
\label{fig:spin_voltage_spin_current}
\end{figure}

Figure \ref{fig:spin_voltage_spin_current} shows the spatial profile of ${\boldsymbol\mu}_{x}$ and ${\boldsymbol j}_{x}$ for several different system parameters, after substituting the ${\boldsymbol\mu}_{0}$ and ${\boldsymbol j}_{0}$ solved self-consistently from Eq.~(\ref{mu0_j0_self_consistent}) into Eq.~(\ref{spin_diffusion_equation}). On the FMM side, we convert the spin expectation value $\langle{\boldsymbol\sigma}_{x}\rangle$ into a spin voltage by ${\boldsymbol\mu}_{x}=a^{3}\langle{\boldsymbol\sigma}_{x}\rangle/N_{F}$ such that the spatial profile of the spin voltage in the entire NM/FMM bilayer can be investigated. The local spin current is calculated from Eq.~(\ref{j0p_j0m}). The spin voltage ${\boldsymbol\mu}_{x}$ in the NM is a smooth function, a result of solving the spin diffusion equation\cite{Zhang00,Chen13}, whereas in the FMM it shows clear feature of oscillation due to the quantum well state described by the $\psi_{F}$ in Eq.~(\ref{NM_FMM_wave_fn}). The precise shape of ${\boldsymbol\mu}_{x}$ in the FMM is a combined effect of quantum interference of the spin transport\cite{Chen15_STT,Ok17} and the self-consistency of Eq.~(\ref{mu0_j0_self_consistent}), and varies significantly with $l_{F}$ and $J_{sd}$, as can be seen by comparing plots with different parameters in Fig.~\ref{fig:spin_voltage_spin_current}. The spin voltage is not continuous across the interface, a result of treating the NM as purely diffusive and the FMM as purely quantum. Rather, the conserved quantity is the spin current according to Eq.~(\ref{j0p_j0m})) is taken into account. As expected, the bigger is the SOC, the more discontinuous is the spin current at the interface, as can be seen by comparing Fig.~\ref{fig:spin_voltage_spin_current} (b) and (d).

\begin{figure}
\centering
\includegraphics[clip=true,width=0.7\columnwidth]{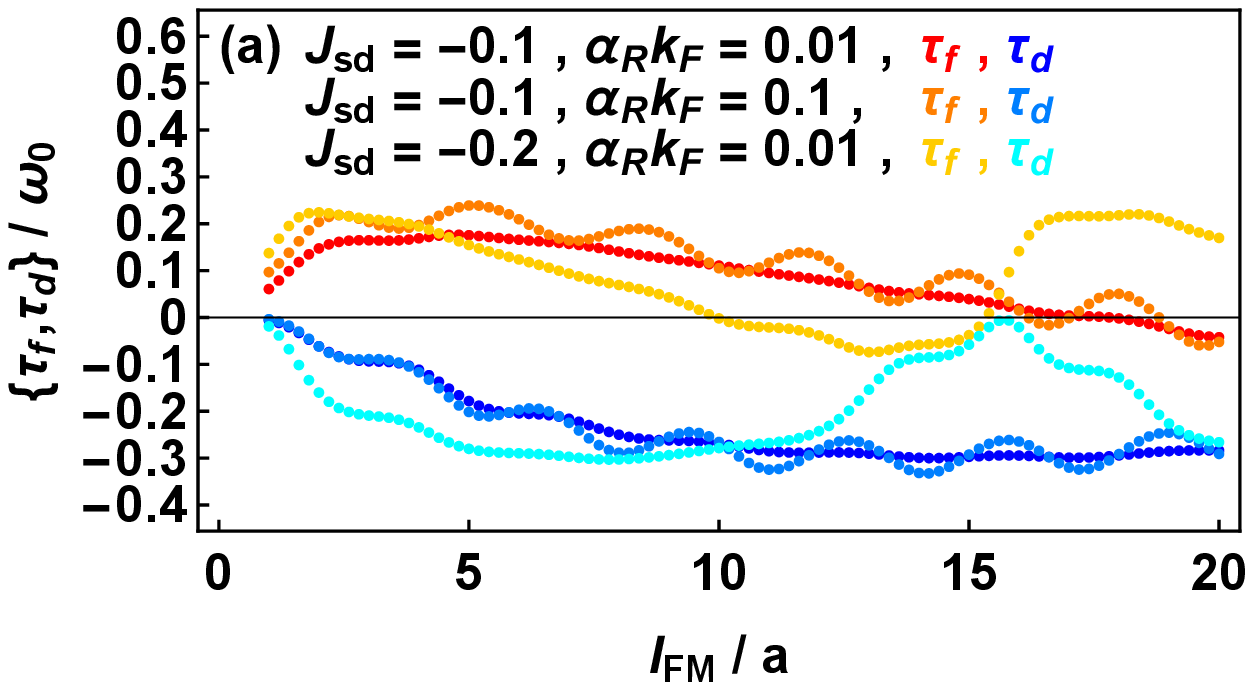}
\includegraphics[clip=true,width=0.7\columnwidth]{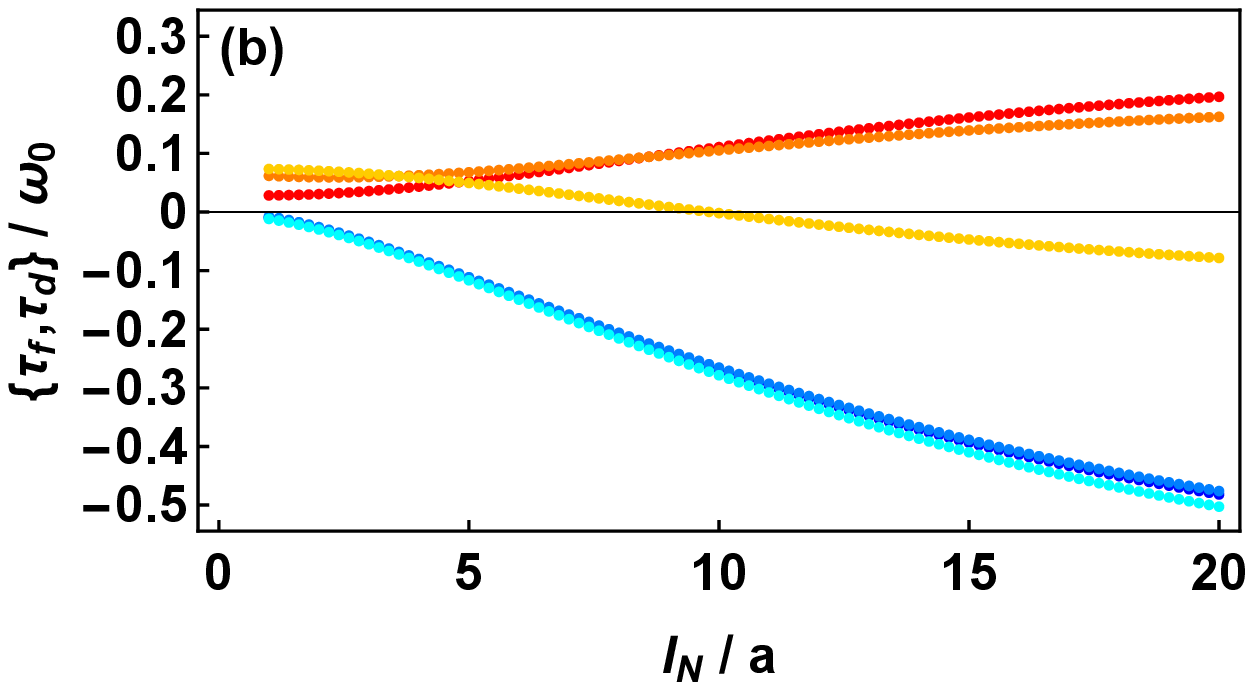}
\includegraphics[clip=true,width=0.7\columnwidth]{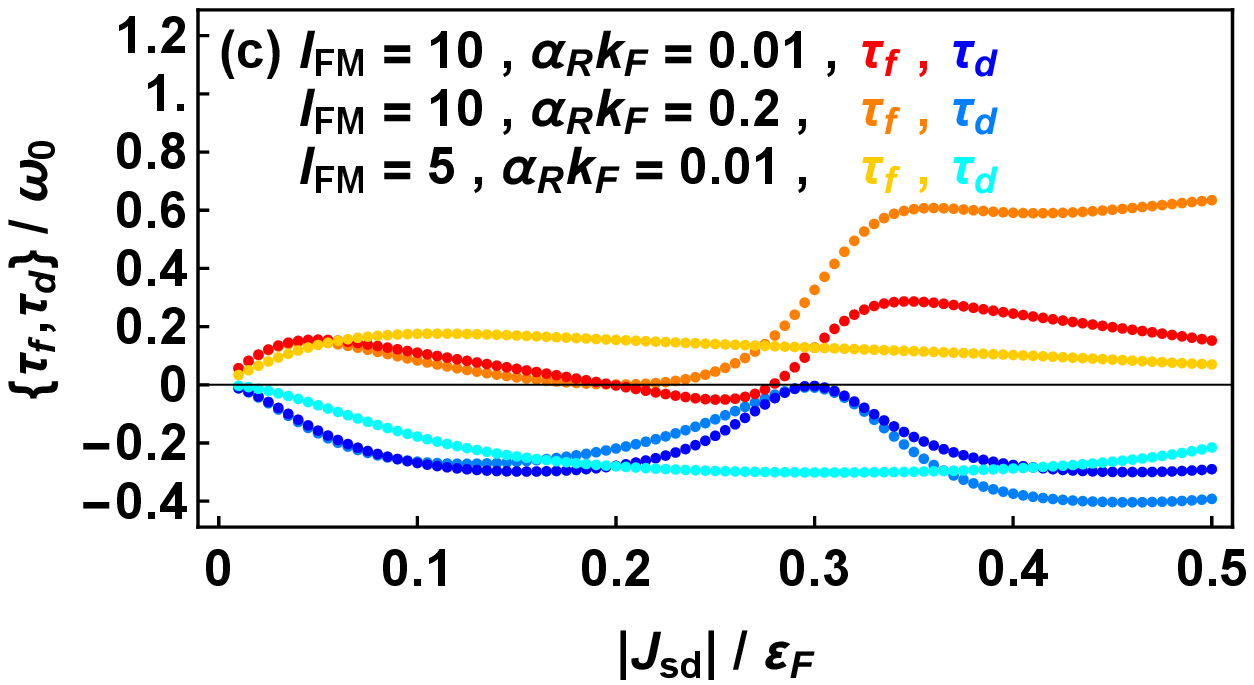}
\includegraphics[clip=true,width=0.7\columnwidth]{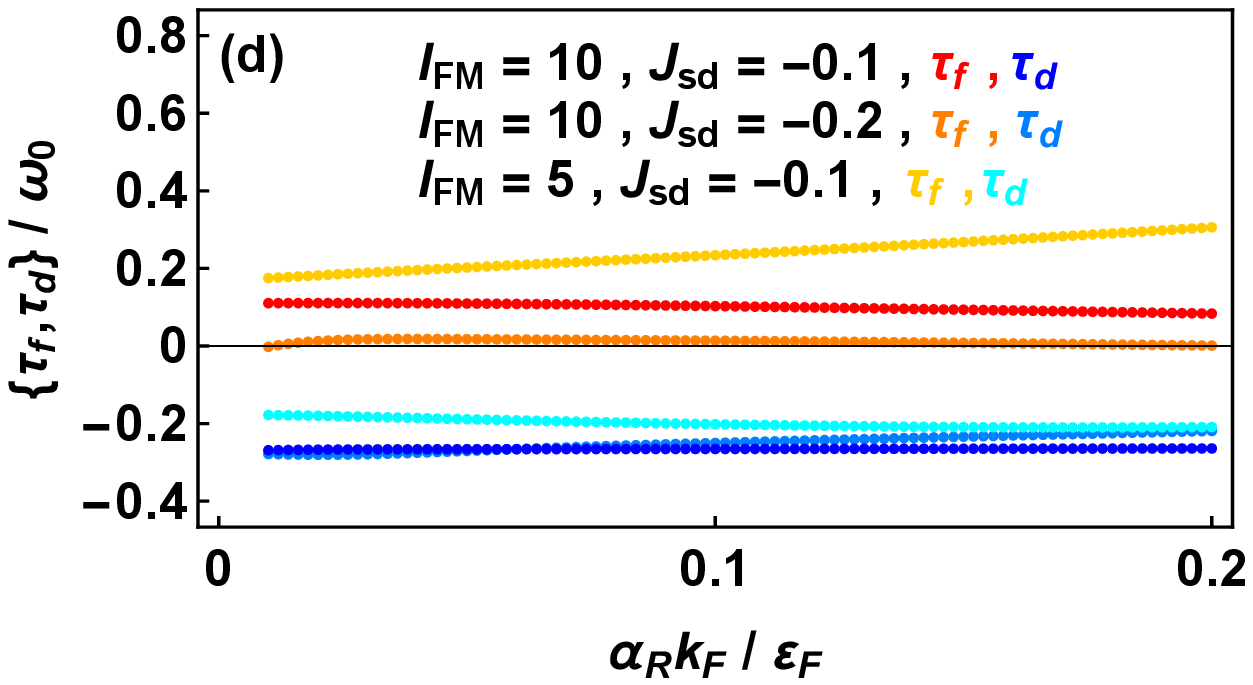}
\caption{ (Color online) Field-like and damping-like $\left\{\tau_{f},\tau_{d}\right\}$ component of the total spin torque versus parameters $\left\{l_{FM},l_{N},J_{sd},\alpha_{R}k_{F}\right\}$ at charge current $j_{c}\sim 10^{11}$A/m$^{2}$, in units of the characteristic frequency $\omega_{0}\sim\omega_{1}\sim 10$ GHz, or equivalently the corresponding magnetic fields $\left\{B_{f},B_{d}\right\}$ in units of $\hbar\omega_{0}/\mu_{B}\sim\hbar\omega_{1}/\mu_{B}\sim 100$mT. The torques versus $l_{FM}$ with fixed $l_{N}=10a$ is shown in (a), and the torques versus $l_{N}$ with fixed $l_{FM}=10a$ is shown in (b). The colors label the same parameters in these two plots. (c) and (d) show the torques versus exchange coupling $J_{sd}$ and interface SOC $\alpha_{R}k_{F}$, respectively, with thickness NM fixed at $l_{N}=10a$. } 
\label{fig:torque_scanlF}
\end{figure}

Figure \ref{fig:torque_scanlF} (a) and (b) shows the thickness dependence of $\tau_{f}$ and $\tau_{d}$. As a function of the FMM thickness $l_{FM}$, both components show a dramatic oscillatory dependence, again due to the spin transport mediated by the quantum well state, which seems to coincide with the experimentally observed rapid variation of the spin torque\cite{Kim13} at least in some reasonable parameter regime. The oscillatory behavior also sensitively depends on the exchange coupling $J_{sd}$. Exceeding certain FMM thickness, the spin torque may change sign. As a function of NM thickness $l_{N}$, both components increase smoothly in the regime $l_{N}\sim \lambda$, owing to the fact that the interface spin voltage ${\boldsymbol\mu}_{0}$ generally increases with $l_{N}$ in this regime, causing an increase in the spin injection and hence the spin torque. Note that the spin torque may change sign with increasing $l_{N}$, which can be a possible mechanism for the reversed domain wall motion observed in Pt/Co/Pt trilayer\cite{Lavrijsen12} with changing Pt thickness (although the quantum interference effect in the trilayer may be even more complicated). The above features all come from the SHE-STT part of the spin torque but not the SOT part, since in our sharp interface approximation the SOT is strictly confined to the interface and does not vary with either $l_{N}$ or $l_{FM}$. The absolute magnitude of the spin torque at current density $j_{c}\sim 10^{11}$A/m$^{2}$, after converting to the effective field, is of the order of few tenth of $\hbar\omega_{0}/\mu_{B}\sim$100mT, close to that observed experimentally\cite{Miron10,Miron11,Garello13}.

The spin torque also shows clear modulation with the exchange coupling $J_{sd}$, as can be seen in Fig.~\ref{fig:torque_scanlF} (c). The oscillatory behavior with respect to both $l_{FM}$ and $J_{sd}$ points to a simple physical picture for how the spin transport mediated by quantum well state yields the SHE-STT: The conduction electron spin injected from the NM precesses around the magnetization when it travels inside the FMM. The phase difference between the injected and the reflected spin yields the SHE-STT. The phase difference is determined by how fast the spin precesses ($J_{sd}/\hbar$) and how much distance it travels ($l_{FM}$), hence the oscillatory behavior with respect to both $J_{sd}$ and $l_{FM}$. We remark that the SOT part also depends on $J_{sd}$, since it enters the Boltzmann of calculating the spin accumulation, yet the dependence is rather smooth, so the rapid variations mainly come from the SHE-STT part. Our result indicates that, even in a clean FMM without other complications like domains or impurities, the quantum tunneling alone can already account for the rapid variation of the spin torque\cite{Kim13}. Without invoking other mechanisms such as spin relaxation\cite{Kim12,Wang12} or spin dependent scattering\cite{Pesin12}, the SOT alone is predominantly field-like at any $\left\{J_{sd},\alpha_{R}k_{F}\right\}$, thus the damping-like components in Fig.~\ref{fig:torque_scanlF} (a) to (d) are almost entirely contributed from the SHE-STT. 

\begin{figure}
\centering
\includegraphics[clip=true,width=0.95\columnwidth]{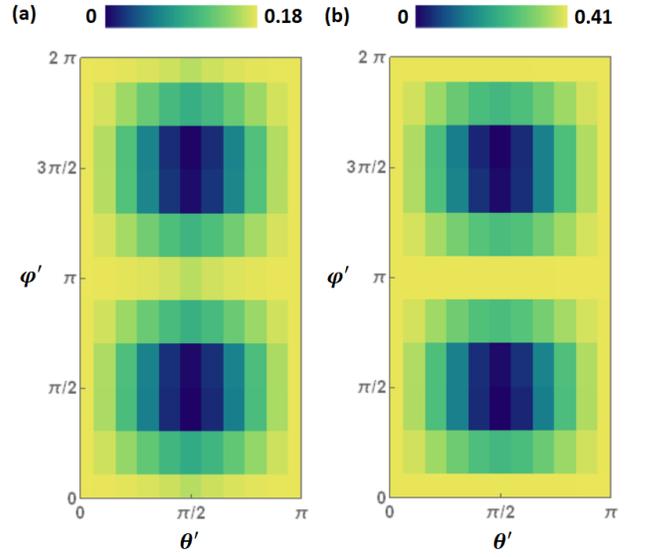}
\caption{ (Color online) The dependence of (a) the field-like torque $\tau_{f}$ and (b) negative of the damping-like torque $-\tau_{d}$ on the angle of magnetization $(\theta^{\prime},\phi^{\prime})$ defined in the experimental coordinate (see Fig.~\ref{fig:NMFMM_schematics}). The parameters are $J_{sd}/\epsilon_{F}=-0.1$, $\alpha_{R}k_{F}=0.01$, $l_{N}/a=10$, $l_{FM}/a=10$, and plots are in the same unit as Fig.~\ref{fig:torque_scanlF}. The two plots show practically the same angular dependence, which are fairly independent from the parameters. } 
\label{fig:tau_scanthetapphip}
\end{figure}

Finally, we remark on the dependence of the spin torque on the direction of the magnetization. Fig.~\ref{fig:tau_scanthetapphip} shows $\tau_{f}$ and $\tau_{d}$ versus the angle $(\theta^{\prime},\varphi^{\prime})$ defined in the experimental coordinate (see Fig.~\ref{fig:NMFMM_schematics}). We use $(\theta^{\prime},\varphi^{\prime})$ such that comparison with experiments can be easily made\cite{Garello13,Qiu14}. The torques are smallest when the magnetization points in-plane and perpendicular to the charge current, as indicated by the blue regions in Fig.~\ref{fig:tau_scanthetapphip}, and increases when the magnetization moves away from this direction. These results are, however, at odds with those revealed by experiments, which concluded that the torque is larger when the magnetization has more in-plane components\cite{Garello13,Qiu14}. Based these results, we speculate that the mechanisms not taken into account by our treatment, such as the Berry phase effect\cite{Lee15} or anisotropic spin relaxation rate\cite{Pauyac13}, may be crucial to explain the angular dependence revealed by experiments.

\section{Conclusions \label{sec:Conclusions}}

In summary, using a semiclassical approach that simultaneously incorporates SHE, spin diffusion, quantum well state, and interface SOC, we address the in-plane current induced spin torque in NM/FMM bilayers. Treating the NM as purely diffusive and the FMM as purely quantum, as considered relevant to most of the experiments, the approach reveals the following intriguing features. Firstly, from the spin diffusion equation one sees that the spatial profile of SHE induced spin voltage is determined by the interface spin current\cite{Chen13}, yet the interface spin current in turn also depends on the spin voltage. Thus, the two have to be solved in a self-consistent manner. Secondly, once the interface spin current is determined, one may naively identify it with the SHE-STT according to the angular momentum conservation\cite{Slonczewski96}. The interface SOC, however, yields an effective interface magnetic field that alters this direct identification, and the interface spin current is equal to the SHE-STT only after the contribution from SOC is subtracted.

Thirdly, assuming the Rashba SOC only exists in the interface atomic layer, the SOT is independent from the thickness of the NM and the FMM, and is not too much influenced by the interface spin voltage. On the other hand, owing to the quantum interference effect of the injected conduction electron, the SHE-STT strongly depends on and may change sign with the thickness of the NM and the FMM, possibly explaining the experimentally observed rapid variation of the spin torque\cite{Kim13} and the sign change of current induced domain wall motion\cite{Lavrijsen12} with respect to the layer thickness. Finally, using the parameters relevant to realistic thin films, the SHE-STT and SOT are revealed to be of the same order of magnitude ($\sim$GHz), comparable to that unveiled by experiments. Besides helping to understand the role of these complicated intertwined ingredients, we anticipate that our approach can help to guide the engineering of the spin torque, whether one aims at changing the magnitude of the spin torque or the relative weighting between its damping-like and field-like components.

\section{Acknowledgement}

We thank exclusively P. Gambardella for the constant input of experimental results, and suggestions leading to refinement of our models. M.~S. and W.~C. acknowledge financial support from a grant of the Swiss National Science Foundation

\appendix

\section{Tunneling amplitudes\label{append:tunneling_amplitudes}}

Upon matching the boundary conditions in Eq.~(\ref{wave_fn_boundary_condition}), and use the short-hand notation
\begin{eqnarray}
\mu_{1}&=&e^{-i \varphi_{\mu} /2} \cos \frac{\theta_{\mu}}{2}\;,\;\;\;\mu_{2}=e^{i \varphi_{\mu} /2} \sin \frac{\theta_{\mu}}{2}\;,
\nonumber \\
\mu_{3}&=&-e^{-i \varphi_{\mu} /2} \sin \frac{\theta_{\mu}}{2}\;,\;\;\;
\mu_{4}=e^{i \varphi_{\mu} /2} \cos \frac{\theta_{\mu}}{2}\;,
\nonumber \\
s_{1}&=&e^{-i \varphi /2} \cos \frac{\theta}{2}\;,\;\;\;s_{2}=e^{i \varphi /2} \sin \frac{\theta}{2}\;,
\nonumber \\
s_{3}&=&-e^{-i \varphi /2} \sin \frac{\theta}{2}\;,\;\;\;
s_{4}=e^{i \varphi /2} \cos \frac{\theta}{2}\;,
\nonumber \\
\lambda_{\alpha\sigma\gamma}&=&ik_{\alpha}\left(1+e^{2ik_{\alpha}l_{FM}}\right)
+\left(ik_{0\sigma}+\gamma\beta_{R}\right)\left(1-e^{2ik_{\alpha}l_{FM}}\right)\;,
\nonumber \\
\xi&=&(\lambda_{+\downarrow-}s_{1}\mu_{2}-\lambda_{+\downarrow+}s_{2}\mu_{1})/(\lambda_{-\downarrow+}s_{4}\mu_{1}
-\lambda_{-\downarrow-}s_{3}\mu_{2})\;,
\nonumber \\
\Omega&=&\lambda_{+\uparrow-}s_{1}\mu_{4}+\lambda_{-\uparrow-}s_{3}\mu_{4}\xi-\lambda_{+\uparrow+}s_{2}\mu_{3}-\lambda_{-\uparrow+}s_{4}\mu_{3}\xi\;,
\nonumber \\
\overline{\Omega}&=&\lambda_{+\uparrow+}s_{2}\mu_{1}+\lambda_{-\uparrow+}s_{4}\mu_{1}\xi-\lambda_{+\uparrow-}s_{1}\mu_{2}-\lambda_{-\uparrow-}s_{3}\mu_{2}\xi\;,
\nonumber \\
\end{eqnarray}
where $\left\{\alpha,\gamma\right\}=\left\{+,-\right\}$ and $\sigma=\left\{\uparrow,\downarrow\right\}$, the scattering coefficients are
\begin{eqnarray}
\frac{B}{A}&=&\frac{k_{0\uparrow}+k_{0\downarrow}}{k_{0\uparrow}-k_{0\downarrow}}
-\frac{2k_{0\uparrow}(\mu_{1}\mu_{4}-\mu_{2}\mu_{3})(\lambda_{+\downarrow-}s_{1}+\lambda_{-\downarrow-}s_{3}\xi)}{(k_{0\uparrow}-k_{0\downarrow})\Omega\mu_{1}}\;,
\nonumber \\
\frac{C}{A}&=&\frac{2k_{0\uparrow}\overline{\Omega}}{(k_{0\uparrow}-k_{0\downarrow})\Omega}\;,
\nonumber \\
\frac{D}{A}&=&2ik_{0\uparrow}\frac{(\mu_{1}\mu_{4}-\mu_{2}\mu_{3})}{\Omega}\;,\;\;\;E=\xi D\;.
\end{eqnarray}
The calculation of spin expectation values and the spin current are then straight forward as described in Sec.~\ref{sec:GrGi_theory}.

\section{Details of calculating the spin accumulation due to SOC\label{append:Boltzmann_eq}}

We provide a formalism that is convenient for numerically calculating the SOC induced spin accumulation in the parabolic band approximation at any $\left\{\alpha_{R}k_{F},J_{sd}\right\}$. The Rashba SOC and exchange coupling in Eq.~(\ref{interface_Hamiltonian}) together constitute an effective, momentum dependent magnetic field in the interface Hamiltonian $H_{I,yz}=\hbar^2 k^2/2m+{\bf B}_{\bf k} \cdot {\boldsymbol \sigma}$, with
\begin{eqnarray}
&&\left(\begin{array}{c}
B_{\bf k}^x \\
B_{\bf k}^y \\
B_{\bf k}^z
\end{array}\right) = \left(\begin{array}{l}
J_{sd}S \sin\theta\cos\varphi  \\
J_{sd} \sin\theta\sin\varphi+\alpha_R k^z \\
J_{sd}S \cos \theta- \alpha_R k^y
\end{array}\right),
\nonumber \\
&&|{\bf B}_{\bf k}|= \left[J_{sd}^2S^{2}+\alpha_R^2 k^2\right. 
\nonumber \\
&&\left.+2 \alpha_R k J_{sd}S (\sin \theta \sin \varphi \sin \xi -\cos \theta \cos \xi)\right]^{1/2}\;,
\end{eqnarray}
and hence the eigenvalues
\begin{equation}
E_{{\bf k},\pm}= \frac{\hbar^2 k^2}{2m} \pm |\bf{B}_k|\;.
\end{equation}
Defining the planar coordinate ${\bf r}=(0,y,z)$, the eigenstates are
\begin{equation}
\psi_{I\pm}(y,z) = \frac{e^{i\bf{k}\cdot\bf{r}}}{a^{3/2}\sqrt{(B_{\bf k}^z \pm |{\bf B}_{\bf k}|)^2 + B_{\perp}^2}}\left(\begin{array}{c}B_{\bf k}^z \pm |{\bf B}_{\bf k}| \\B_{\perp}^{\ast}\end{array}\right)
\end{equation}
where we define $B_{\perp}=B_{\bf k}^{x}-iB_{\bf k}^{y}$. Consequently, the $\alpha$-component spin expectation value of the eigenstate is $\langle{\sigma}^{\alpha}\rangle_{\pm}=\psi_{I\pm}^{\ast}(y,z)\sigma^{\alpha}\psi_{I\pm}(y,z)=\pm B_{\bf k}^\alpha/|{\bf B}_{\bf k}|a^{3}$. The spin accumulation is then, by expressing Eq.~(\ref{Boltzmann_eq}) in polar coordinate,
\begin{eqnarray}
&&\langle\delta{\boldsymbol \sigma}\rangle_{\pm}=\frac{a^{2}e\tau E_y}{4\pi^2 \hbar}\int k\,dk\,d\xi\langle{\boldsymbol \sigma}\rangle_{\pm}
\nonumber \\
&&\times\left(\cos \xi \frac{\partial E_{{\bf k,\pm}}}{\partial k}-\frac{1}{k}\sin \xi \frac{\partial E_{{\bf k,\pm}}}{\partial \xi}\right)
\left(-\frac{\partial f_{{\bf k},\pm}^{0}}{\partial  E_{{\bf k,\pm}}}\right)\;.
\end{eqnarray}
Numerically, by defining the square root $q_{i}$ of various energy scales 
\begin{eqnarray}
&&\frac{\hbar^{2}k^{2}}{2m}=q^{2}\;,\;\;\;\frac{\hbar^{2}k_{F}^{2}}{2m}=q_{F}^{2}\;,
\;\;\;k_{B}T=q_{T}^{2}\;,
\nonumber \\
&&\alpha_{R}k=q_{\alpha}q\;,\;\;\;J_{sd}S=q_{J}^{2}<0\;,
\nonumber \\
&&|{\bf B}_{\bf k}|=\left[q_{J}^{4}+q_{\alpha}^{2}q^{2}+2qq_{\alpha}q_{J}^{2}\left(\sin\theta\sin\varphi\sin\xi\right.\right.
\nonumber \\
&&\left.\left.-\cos\theta\cos\xi\right)\right]^{1/2}\;,
\end{eqnarray}
we can evaluate the integral at any $\left\{\alpha_{R}k_{F},J_{sd},\theta,\varphi\right\}$ by explicitly calculating
\begin{eqnarray}
&&\langle\delta {\boldsymbol \sigma}\rangle_{\pm} = \left(\frac{e}{4\pi^{2}\hbar a}\tau E\frac{\sqrt{2m}}{\hbar}\right)\int q\; dq\; d\xi 
\nonumber \\
  &&\times \left[q\cos \xi \left(2\pm \frac{q_{\alpha}^2}{|B_k|}\right)\mp \frac{q_{\alpha}g_{J}^2 \cos\theta}{|B_k|}\right]
\nonumber \\
&&\times (\pm)
\left(\begin{array}{c}
q_{J}^{2}\sin\theta\cos\varphi \\
q_{J}^{2}\sin\theta\sin\varphi+q_{\alpha}q\sin\xi \\
q_{J}^{2}\cos\theta-q_{\alpha}q\cos\xi 
\end{array}\right) 
\nonumber \\
  &&\times\frac{1}{|{\bf B}_{\bf k}|q_T^2}\times\frac{e^{(q^2\pm |{\bf B}_{\bf k}|-q_F^2)/q_T^2}}{[e^{(q^2 \pm |{\bf B}_{\bf k}| -q_F^2)/q_T^2}+1]^2}\;.
\label{deltasigma_integral}
\end{eqnarray}
For all the simulation presented in Sec.~\ref{sec:numerical_results}, we fix the temperature at $k_{B}T/\epsilon_{F}=0.02$. For any $\left\{\alpha_{R}k_{F},J_{sd}\right\}$, $\langle\delta {\boldsymbol \sigma}\rangle_{\pm}$ is even under the inversion of magnetization ${\boldsymbol S}\rightarrow-{\boldsymbol S}$, in agreement with several analytical limits reported previously\cite{Gambardella11,Wang15}. This can be seen by noting that the integral of Eq.~(\ref{deltasigma_integral}) is invariant under $\left\{S^{x},S^{y},S^{z}\right\}\rightarrow\left\{-S^{x},-S^{y},-S^{z}\right\}$ and a shift in the angular argument $\xi\rightarrow\pi+\xi$.

\bibliography{Literatur}

\begin{thebibliography}{51}%
\makeatletter
\providecommand \@ifxundefined [1]{%
 \@ifx{#1\undefined}
}%
\providecommand \@ifnum [1]{%
 \ifnum #1\expandafter \@firstoftwo
 \else \expandafter \@secondoftwo
 \fi
}%
\providecommand \@ifx [1]{%
 \ifx #1\expandafter \@firstoftwo
 \else \expandafter \@secondoftwo
 \fi
}%
\providecommand \natexlab [1]{#1}%
\providecommand \enquote  [1]{``#1''}%
\providecommand \bibnamefont  [1]{#1}%
\providecommand \bibfnamefont [1]{#1}%
\providecommand \citenamefont [1]{#1}%
\providecommand \href@noop [0]{\@secondoftwo}%
\providecommand \href [0]{\begingroup \@sanitize@url \@href}%
\providecommand \@href[1]{\@@startlink{#1}\@@href}%
\providecommand \@@href[1]{\endgroup#1\@@endlink}%
\providecommand \@sanitize@url [0]{\catcode `\\12\catcode `\$12\catcode
  `\&12\catcode `\#12\catcode `\^12\catcode `\_12\catcode `\%12\relax}%
\providecommand \@@startlink[1]{}%
\providecommand \@@endlink[0]{}%
\providecommand \url  [0]{\begingroup\@sanitize@url \@url }%
\providecommand \@url [1]{\endgroup\@href {#1}{\urlprefix }}%
\providecommand \urlprefix  [0]{URL }%
\providecommand \Eprint [0]{\href }%
\providecommand \doibase [0]{http://dx.doi.org/}%
\providecommand \selectlanguage [0]{\@gobble}%
\providecommand \bibinfo  [0]{\@secondoftwo}%
\providecommand \bibfield  [0]{\@secondoftwo}%
\providecommand \translation [1]{[#1]}%
\providecommand \BibitemOpen [0]{}%
\providecommand \bibitemStop [0]{}%
\providecommand \bibitemNoStop [0]{.\EOS\space}%
\providecommand \EOS [0]{\spacefactor3000\relax}%
\providecommand \BibitemShut  [1]{\csname bibitem#1\endcsname}%
\let\auto@bib@innerbib\@empty
\bibitem [{\citenamefont {Berger}(1996)}]{Berger96}%
  \BibitemOpen
  \bibfield  {author} {\bibinfo {author} {\bibfnamefont {L.}~\bibnamefont
  {Berger}},\ }\href {\doibase 10.1103/PhysRevB.54.9353} {\bibfield  {journal}
  {\bibinfo  {journal} {Phys. Rev. B}\ }\textbf {\bibinfo {volume} {54}},\
  \bibinfo {pages} {9353} (\bibinfo {year} {1996})}\BibitemShut {NoStop}%
\bibitem [{\citenamefont {Slonczewski}(1996)}]{Slonczewski96}%
  \BibitemOpen
  \bibfield  {author} {\bibinfo {author} {\bibfnamefont {J.}~\bibnamefont
  {Slonczewski}},\ }\href {\doibase
  http://dx.doi.org/10.1016/0304-8853(96)00062-5} {\bibfield  {journal}
  {\bibinfo  {journal} {J. Magn. Magn. Mater.}\ }\textbf {\bibinfo {volume}
  {159}},\ \bibinfo {pages} {L1 } (\bibinfo {year} {1996})}\BibitemShut
  {NoStop}%
\bibitem [{\citenamefont {Stiles}\ and\ \citenamefont
  {Zangwill}(2002)}]{Stiles02}%
  \BibitemOpen
  \bibfield  {author} {\bibinfo {author} {\bibfnamefont {M.~D.}\ \bibnamefont
  {Stiles}}\ and\ \bibinfo {author} {\bibfnamefont {A.}~\bibnamefont
  {Zangwill}},\ }\href {\doibase 10.1103/PhysRevB.66.014407} {\bibfield
  {journal} {\bibinfo  {journal} {Phys. Rev. B}\ }\textbf {\bibinfo {volume}
  {66}},\ \bibinfo {pages} {014407} (\bibinfo {year} {2002})}\BibitemShut
  {NoStop}%
\bibitem [{\citenamefont {Zhang}\ \emph {et~al.}(2002)\citenamefont {Zhang},
  \citenamefont {Levy},\ and\ \citenamefont {Fert}}]{Zhang02}%
  \BibitemOpen
  \bibfield  {author} {\bibinfo {author} {\bibfnamefont {S.}~\bibnamefont
  {Zhang}}, \bibinfo {author} {\bibfnamefont {P.~M.}\ \bibnamefont {Levy}}, \
  and\ \bibinfo {author} {\bibfnamefont {A.}~\bibnamefont {Fert}},\ }\href
  {\doibase 10.1103/PhysRevLett.88.236601} {\bibfield  {journal} {\bibinfo
  {journal} {Phys. Rev. Lett.}\ }\textbf {\bibinfo {volume} {88}},\ \bibinfo
  {pages} {236601} (\bibinfo {year} {2002})}\BibitemShut {NoStop}%
\bibitem [{\citenamefont {Stiles}\ and\ \citenamefont
  {Miltat}(2006)}]{Stiles06}%
  \BibitemOpen
  \bibfield  {author} {\bibinfo {author} {\bibfnamefont {M.~D.}\ \bibnamefont
  {Stiles}}\ and\ \bibinfo {author} {\bibfnamefont {J.}~\bibnamefont
  {Miltat}},\ }\enquote {\bibinfo {title} {Spin-transfer torque and
  dynamics},}\ in\ \href {\doibase 10.1007/10938171_7} {\emph {\bibinfo
  {booktitle} {Spin Dynamics in Confined Magnetic Structures III}}},\ \bibinfo
  {editor} {edited by\ \bibinfo {editor} {\bibfnamefont {B.}~\bibnamefont
  {Hillebrands}}\ and\ \bibinfo {editor} {\bibfnamefont {A.}~\bibnamefont
  {Thiaville}}}\ (\bibinfo  {publisher} {Springer Berlin Heidelberg},\ \bibinfo
  {address} {Berlin, Heidelberg},\ \bibinfo {year} {2006})\ pp.\ \bibinfo
  {pages} {225--308}\BibitemShut {NoStop}%
\bibitem [{\citenamefont {Ralph}\ and\ \citenamefont {Stiles}(2008)}]{Ralph08}%
  \BibitemOpen
  \bibfield  {author} {\bibinfo {author} {\bibfnamefont {D.}~\bibnamefont
  {Ralph}}\ and\ \bibinfo {author} {\bibfnamefont {M.}~\bibnamefont {Stiles}},\
  }\href {\doibase http://dx.doi.org/10.1016/j.jmmm.2007.12.019} {\bibfield
  {journal} {\bibinfo  {journal} {J. Magn. Magn. Mater.}\ }\textbf {\bibinfo
  {volume} {320}},\ \bibinfo {pages} {1190 } (\bibinfo {year}
  {2008})}\BibitemShut {NoStop}%
\bibitem [{\citenamefont {Miron}\ \emph {et~al.}(2010)\citenamefont {Miron},
  \citenamefont {Gaudin}, \citenamefont {Auffret}, \citenamefont {Rodmacq},
  \citenamefont {Schuhl}, \citenamefont {Pizzini}, \citenamefont {Vogel},\ and\
  \citenamefont {Gambardella}}]{Miron10}%
  \BibitemOpen
  \bibfield  {author} {\bibinfo {author} {\bibfnamefont {I.~M.}\ \bibnamefont
  {Miron}}, \bibinfo {author} {\bibfnamefont {G.}~\bibnamefont {Gaudin}},
  \bibinfo {author} {\bibfnamefont {S.}~\bibnamefont {Auffret}}, \bibinfo
  {author} {\bibfnamefont {B.}~\bibnamefont {Rodmacq}}, \bibinfo {author}
  {\bibfnamefont {A.}~\bibnamefont {Schuhl}}, \bibinfo {author} {\bibfnamefont
  {S.}~\bibnamefont {Pizzini}}, \bibinfo {author} {\bibfnamefont
  {J.}~\bibnamefont {Vogel}}, \ and\ \bibinfo {author} {\bibfnamefont
  {P.}~\bibnamefont {Gambardella}},\ }\href {\doibase 10.1038/nmat2613}
  {\bibfield  {journal} {\bibinfo  {journal} {Nat. Mater.}\ }\textbf {\bibinfo
  {volume} {9}},\ \bibinfo {pages} {230} (\bibinfo {year} {2010})}\BibitemShut
  {NoStop}%
\bibitem [{\citenamefont {Miron}\ \emph
  {et~al.}(2011{\natexlab{a}})\citenamefont {Miron}, \citenamefont {Garello},
  \citenamefont {Gaudin}, \citenamefont {Zermatten}, \citenamefont {Costache},
  \citenamefont {Auffret}, \citenamefont {Bandiera}, \citenamefont {Rodmacq},
  \citenamefont {Schuhl},\ and\ \citenamefont {Gambardella}}]{Miron11}%
  \BibitemOpen
  \bibfield  {author} {\bibinfo {author} {\bibfnamefont {I.~M.}\ \bibnamefont
  {Miron}}, \bibinfo {author} {\bibfnamefont {K.}~\bibnamefont {Garello}},
  \bibinfo {author} {\bibfnamefont {G.}~\bibnamefont {Gaudin}}, \bibinfo
  {author} {\bibfnamefont {P.-J.}\ \bibnamefont {Zermatten}}, \bibinfo {author}
  {\bibfnamefont {M.~V.}\ \bibnamefont {Costache}}, \bibinfo {author}
  {\bibfnamefont {S.}~\bibnamefont {Auffret}}, \bibinfo {author} {\bibfnamefont
  {S.}~\bibnamefont {Bandiera}}, \bibinfo {author} {\bibfnamefont
  {B.}~\bibnamefont {Rodmacq}}, \bibinfo {author} {\bibfnamefont
  {A.}~\bibnamefont {Schuhl}}, \ and\ \bibinfo {author} {\bibfnamefont
  {P.}~\bibnamefont {Gambardella}},\ }\href {\doibase 10.1038/nature10309}
  {\bibfield  {journal} {\bibinfo  {journal} {Nature}\ }\textbf {\bibinfo
  {volume} {476}},\ \bibinfo {pages} {189} (\bibinfo {year}
  {2011}{\natexlab{a}})}\BibitemShut {NoStop}%
\bibitem [{\citenamefont {Liu}\ \emph {et~al.}(2012{\natexlab{a}})\citenamefont
  {Liu}, \citenamefont {Pai}, \citenamefont {Li}, \citenamefont {Tseng},
  \citenamefont {Ralph},\ and\ \citenamefont {Buhrman}}]{Liu12}%
  \BibitemOpen
  \bibfield  {author} {\bibinfo {author} {\bibfnamefont {L.}~\bibnamefont
  {Liu}}, \bibinfo {author} {\bibfnamefont {C.-F.}\ \bibnamefont {Pai}},
  \bibinfo {author} {\bibfnamefont {Y.}~\bibnamefont {Li}}, \bibinfo {author}
  {\bibfnamefont {H.~W.}\ \bibnamefont {Tseng}}, \bibinfo {author}
  {\bibfnamefont {D.~C.}\ \bibnamefont {Ralph}}, \ and\ \bibinfo {author}
  {\bibfnamefont {R.~A.}\ \bibnamefont {Buhrman}},\ }\href {\doibase
  10.1126/science.1218197} {\bibfield  {journal} {\bibinfo  {journal}
  {Science}\ }\textbf {\bibinfo {volume} {336}},\ \bibinfo {pages} {555}
  (\bibinfo {year} {2012}{\natexlab{a}})}\BibitemShut {NoStop}%
\bibitem [{\citenamefont {Liu}\ \emph {et~al.}(2012{\natexlab{b}})\citenamefont
  {Liu}, \citenamefont {Lee}, \citenamefont {Gudmundsen}, \citenamefont
  {Ralph},\ and\ \citenamefont {Buhrman}}]{Liu12_2}%
  \BibitemOpen
  \bibfield  {author} {\bibinfo {author} {\bibfnamefont {L.}~\bibnamefont
  {Liu}}, \bibinfo {author} {\bibfnamefont {O.~J.}\ \bibnamefont {Lee}},
  \bibinfo {author} {\bibfnamefont {T.~J.}\ \bibnamefont {Gudmundsen}},
  \bibinfo {author} {\bibfnamefont {D.~C.}\ \bibnamefont {Ralph}}, \ and\
  \bibinfo {author} {\bibfnamefont {R.~A.}\ \bibnamefont {Buhrman}},\ }\href
  {\doibase 10.1103/PhysRevLett.109.096602} {\bibfield  {journal} {\bibinfo
  {journal} {Phys. Rev. Lett.}\ }\textbf {\bibinfo {volume} {109}},\ \bibinfo
  {pages} {096602} (\bibinfo {year} {2012}{\natexlab{b}})}\BibitemShut
  {NoStop}%
\bibitem [{\citenamefont {Garello}\ \emph {et~al.}(2013)\citenamefont
  {Garello}, \citenamefont {Miron}, \citenamefont {Avci}, \citenamefont
  {Freimuth}, \citenamefont {Mokrousov}, \citenamefont {Blugel}, \citenamefont
  {Auffret}, \citenamefont {Boulle}, \citenamefont {Gaudin},\ and\
  \citenamefont {Gambardella}}]{Garello13}%
  \BibitemOpen
  \bibfield  {author} {\bibinfo {author} {\bibfnamefont {K.}~\bibnamefont
  {Garello}}, \bibinfo {author} {\bibfnamefont {I.~M.}\ \bibnamefont {Miron}},
  \bibinfo {author} {\bibfnamefont {C.~O.}\ \bibnamefont {Avci}}, \bibinfo
  {author} {\bibfnamefont {F.}~\bibnamefont {Freimuth}}, \bibinfo {author}
  {\bibfnamefont {Y.}~\bibnamefont {Mokrousov}}, \bibinfo {author}
  {\bibfnamefont {S.}~\bibnamefont {Blugel}}, \bibinfo {author} {\bibfnamefont
  {S.}~\bibnamefont {Auffret}}, \bibinfo {author} {\bibfnamefont
  {O.}~\bibnamefont {Boulle}}, \bibinfo {author} {\bibfnamefont
  {G.}~\bibnamefont {Gaudin}}, \ and\ \bibinfo {author} {\bibfnamefont
  {P.}~\bibnamefont {Gambardella}},\ }\href {\doibase 10.1038/nnano.2013.145}
  {\bibfield  {journal} {\bibinfo  {journal} {Nat. Nano.}\ }\textbf {\bibinfo
  {volume} {8}},\ \bibinfo {pages} {587} (\bibinfo {year} {2013})}\BibitemShut
  {NoStop}%
\bibitem [{\citenamefont {Obata}\ and\ \citenamefont {Tatara}(2008)}]{Obata08}%
  \BibitemOpen
  \bibfield  {author} {\bibinfo {author} {\bibfnamefont {K.}~\bibnamefont
  {Obata}}\ and\ \bibinfo {author} {\bibfnamefont {G.}~\bibnamefont {Tatara}},\
  }\href {\doibase 10.1103/PhysRevB.77.214429} {\bibfield  {journal} {\bibinfo
  {journal} {Phys. Rev. B}\ }\textbf {\bibinfo {volume} {77}},\ \bibinfo
  {pages} {214429} (\bibinfo {year} {2008})}\BibitemShut {NoStop}%
\bibitem [{\citenamefont {Manchon}\ and\ \citenamefont
  {Zhang}(2008)}]{Manchon08}%
  \BibitemOpen
  \bibfield  {author} {\bibinfo {author} {\bibfnamefont {A.}~\bibnamefont
  {Manchon}}\ and\ \bibinfo {author} {\bibfnamefont {S.}~\bibnamefont
  {Zhang}},\ }\href {\doibase 10.1103/PhysRevB.78.212405} {\bibfield  {journal}
  {\bibinfo  {journal} {Phys. Rev. B}\ }\textbf {\bibinfo {volume} {78}},\
  \bibinfo {pages} {212405} (\bibinfo {year} {2008})}\BibitemShut {NoStop}%
\bibitem [{\citenamefont {Manchon}\ and\ \citenamefont
  {Zhang}(2009)}]{Manchon09}%
  \BibitemOpen
  \bibfield  {author} {\bibinfo {author} {\bibfnamefont {A.}~\bibnamefont
  {Manchon}}\ and\ \bibinfo {author} {\bibfnamefont {S.}~\bibnamefont
  {Zhang}},\ }\href {\doibase 10.1103/PhysRevB.79.094422} {\bibfield  {journal}
  {\bibinfo  {journal} {Phys. Rev. B}\ }\textbf {\bibinfo {volume} {79}},\
  \bibinfo {pages} {094422} (\bibinfo {year} {2009})}\BibitemShut {NoStop}%
\bibitem [{\citenamefont {Matos-Abiague}\ and\ \citenamefont
  {Rodr\'{\i}guez-Su\'arez}(2009)}]{Matos-Abiague09}%
  \BibitemOpen
  \bibfield  {author} {\bibinfo {author} {\bibfnamefont {A.}~\bibnamefont
  {Matos-Abiague}}\ and\ \bibinfo {author} {\bibfnamefont {R.~L.}\ \bibnamefont
  {Rodr\'{\i}guez-Su\'arez}},\ }\href {\doibase 10.1103/PhysRevB.80.094424}
  {\bibfield  {journal} {\bibinfo  {journal} {Phys. Rev. B}\ }\textbf {\bibinfo
  {volume} {80}},\ \bibinfo {pages} {094424} (\bibinfo {year}
  {2009})}\BibitemShut {NoStop}%
\bibitem [{\citenamefont {Haney}\ and\ \citenamefont {Stiles}(2010)}]{Haney10}%
  \BibitemOpen
  \bibfield  {author} {\bibinfo {author} {\bibfnamefont {P.~M.}\ \bibnamefont
  {Haney}}\ and\ \bibinfo {author} {\bibfnamefont {M.~D.}\ \bibnamefont
  {Stiles}},\ }\href {\doibase 10.1103/PhysRevLett.105.126602} {\bibfield
  {journal} {\bibinfo  {journal} {Phys. Rev. Lett.}\ }\textbf {\bibinfo
  {volume} {105}},\ \bibinfo {pages} {126602} (\bibinfo {year}
  {2010})}\BibitemShut {NoStop}%
\bibitem [{\citenamefont {Gambardella}\ and\ \citenamefont
  {Miron}(2011)}]{Gambardella11}%
  \BibitemOpen
  \bibfield  {author} {\bibinfo {author} {\bibfnamefont {P.}~\bibnamefont
  {Gambardella}}\ and\ \bibinfo {author} {\bibfnamefont {I.~M.}\ \bibnamefont
  {Miron}},\ }\href {\doibase 10.1098/rsta.2010.0336} {\bibfield  {journal}
  {\bibinfo  {journal} {Phil. Trans. R. Soc. A}\ }\textbf {\bibinfo {volume}
  {369}},\ \bibinfo {pages} {3175} (\bibinfo {year} {2011})}\BibitemShut
  {NoStop}%
\bibitem [{\citenamefont {Hirsch}(1999)}]{Hirsch99}%
  \BibitemOpen
  \bibfield  {author} {\bibinfo {author} {\bibfnamefont {J.~E.}\ \bibnamefont
  {Hirsch}},\ }\href {\doibase 10.1103/PhysRevLett.83.1834} {\bibfield
  {journal} {\bibinfo  {journal} {Phys. Rev. Lett.}\ }\textbf {\bibinfo
  {volume} {83}},\ \bibinfo {pages} {1834} (\bibinfo {year}
  {1999})}\BibitemShut {NoStop}%
\bibitem [{\citenamefont {Murakami}\ \emph {et~al.}(2003)\citenamefont
  {Murakami}, \citenamefont {Nagaosa},\ and\ \citenamefont
  {Zhang}}]{Murakami03}%
  \BibitemOpen
  \bibfield  {author} {\bibinfo {author} {\bibfnamefont {S.}~\bibnamefont
  {Murakami}}, \bibinfo {author} {\bibfnamefont {N.}~\bibnamefont {Nagaosa}}, \
  and\ \bibinfo {author} {\bibfnamefont {S.-C.}\ \bibnamefont {Zhang}},\ }\href
  {\doibase 10.1126/science.1087128} {\bibfield  {journal} {\bibinfo  {journal}
  {Science}\ }\textbf {\bibinfo {volume} {301}},\ \bibinfo {pages} {1348}
  (\bibinfo {year} {2003})}\BibitemShut {NoStop}%
\bibitem [{\citenamefont {Sinova}\ \emph {et~al.}(2004)\citenamefont {Sinova},
  \citenamefont {Culcer}, \citenamefont {Niu}, \citenamefont {Sinitsyn},
  \citenamefont {Jungwirth},\ and\ \citenamefont {MacDonald}}]{Sinova04}%
  \BibitemOpen
  \bibfield  {author} {\bibinfo {author} {\bibfnamefont {J.}~\bibnamefont
  {Sinova}}, \bibinfo {author} {\bibfnamefont {D.}~\bibnamefont {Culcer}},
  \bibinfo {author} {\bibfnamefont {Q.}~\bibnamefont {Niu}}, \bibinfo {author}
  {\bibfnamefont {N.~A.}\ \bibnamefont {Sinitsyn}}, \bibinfo {author}
  {\bibfnamefont {T.}~\bibnamefont {Jungwirth}}, \ and\ \bibinfo {author}
  {\bibfnamefont {A.~H.}\ \bibnamefont {MacDonald}},\ }\href {\doibase
  10.1103/PhysRevLett.92.126603} {\bibfield  {journal} {\bibinfo  {journal}
  {Phys. Rev. Lett.}\ }\textbf {\bibinfo {volume} {92}},\ \bibinfo {pages}
  {126603} (\bibinfo {year} {2004})}\BibitemShut {NoStop}%
\bibitem [{\citenamefont {Sinova}\ \emph {et~al.}(2015)\citenamefont {Sinova},
  \citenamefont {Valenzuela}, \citenamefont {Wunderlich}, \citenamefont
  {Back},\ and\ \citenamefont {Jungwirth}}]{Sinova15}%
  \BibitemOpen
  \bibfield  {author} {\bibinfo {author} {\bibfnamefont {J.}~\bibnamefont
  {Sinova}}, \bibinfo {author} {\bibfnamefont {S.~O.}\ \bibnamefont
  {Valenzuela}}, \bibinfo {author} {\bibfnamefont {J.}~\bibnamefont
  {Wunderlich}}, \bibinfo {author} {\bibfnamefont {C.~H.}\ \bibnamefont
  {Back}}, \ and\ \bibinfo {author} {\bibfnamefont {T.}~\bibnamefont
  {Jungwirth}},\ }\href {\doibase 10.1103/RevModPhys.87.1213} {\bibfield
  {journal} {\bibinfo  {journal} {Rev. Mod. Phys.}\ }\textbf {\bibinfo {volume}
  {87}},\ \bibinfo {pages} {1213} (\bibinfo {year} {2015})}\BibitemShut
  {NoStop}%
\bibitem [{\citenamefont {McGuire}\ and\ \citenamefont
  {Potter}(1975)}]{McGuire75}%
  \BibitemOpen
  \bibfield  {author} {\bibinfo {author} {\bibfnamefont {T.}~\bibnamefont
  {McGuire}}\ and\ \bibinfo {author} {\bibfnamefont {R.}~\bibnamefont
  {Potter}},\ }\href {\doibase 10.1109/TMAG.1975.1058782} {\bibfield  {journal}
  {\bibinfo  {journal} {IEEE Transactions on Magnetics}\ }\textbf {\bibinfo
  {volume} {11}},\ \bibinfo {pages} {1018} (\bibinfo {year}
  {1975})}\BibitemShut {NoStop}%
\bibitem [{\citenamefont {Avci}\ \emph {et~al.}(2015)\citenamefont {Avci},
  \citenamefont {Garello}, \citenamefont {Ghosh}, \citenamefont {Gabureac},
  \citenamefont {Alvarado},\ and\ \citenamefont {Gambardella}}]{Avci15}%
  \BibitemOpen
  \bibfield  {author} {\bibinfo {author} {\bibfnamefont {C.~O.}\ \bibnamefont
  {Avci}}, \bibinfo {author} {\bibfnamefont {K.}~\bibnamefont {Garello}},
  \bibinfo {author} {\bibfnamefont {A.}~\bibnamefont {Ghosh}}, \bibinfo
  {author} {\bibfnamefont {M.}~\bibnamefont {Gabureac}}, \bibinfo {author}
  {\bibfnamefont {S.~F.}\ \bibnamefont {Alvarado}}, \ and\ \bibinfo {author}
  {\bibfnamefont {P.}~\bibnamefont {Gambardella}},\ }\href {\doibase
  10.1038/nphys3356} {\bibfield  {journal} {\bibinfo  {journal} {Nature Phys.}\
  }\textbf {\bibinfo {volume} {11}},\ \bibinfo {pages} {570} (\bibinfo {year}
  {2015})}\BibitemShut {NoStop}%
\bibitem [{\citenamefont {Zhang}\ and\ \citenamefont
  {Vignale}(2016)}]{Zhang17}%
  \BibitemOpen
  \bibfield  {author} {\bibinfo {author} {\bibfnamefont {S.~S.-L.}\
  \bibnamefont {Zhang}}\ and\ \bibinfo {author} {\bibfnamefont
  {G.}~\bibnamefont {Vignale}},\ }\href {\doibase 10.1103/PhysRevB.94.140411}
  {\bibfield  {journal} {\bibinfo  {journal} {Phys. Rev. B}\ }\textbf {\bibinfo
  {volume} {94}},\ \bibinfo {pages} {140411} (\bibinfo {year}
  {2016})}\BibitemShut {NoStop}%
\bibitem [{\citenamefont {Zhang}(2000)}]{Zhang00}%
  \BibitemOpen
  \bibfield  {author} {\bibinfo {author} {\bibfnamefont {S.}~\bibnamefont
  {Zhang}},\ }\href {\doibase 10.1103/PhysRevLett.85.393} {\bibfield  {journal}
  {\bibinfo  {journal} {Phys. Rev. Lett.}\ }\textbf {\bibinfo {volume} {85}},\
  \bibinfo {pages} {393} (\bibinfo {year} {2000})}\BibitemShut {NoStop}%
\bibitem [{\citenamefont {Chen}\ \emph {et~al.}(2013)\citenamefont {Chen},
  \citenamefont {Takahashi}, \citenamefont {Nakayama}, \citenamefont
  {Althammer}, \citenamefont {Goennenwein}, \citenamefont {Saitoh},\ and\
  \citenamefont {Bauer}}]{Chen13}%
  \BibitemOpen
  \bibfield  {author} {\bibinfo {author} {\bibfnamefont {Y.-T.}\ \bibnamefont
  {Chen}}, \bibinfo {author} {\bibfnamefont {S.}~\bibnamefont {Takahashi}},
  \bibinfo {author} {\bibfnamefont {H.}~\bibnamefont {Nakayama}}, \bibinfo
  {author} {\bibfnamefont {M.}~\bibnamefont {Althammer}}, \bibinfo {author}
  {\bibfnamefont {S.~T.~B.}\ \bibnamefont {Goennenwein}}, \bibinfo {author}
  {\bibfnamefont {E.}~\bibnamefont {Saitoh}}, \ and\ \bibinfo {author}
  {\bibfnamefont {G.~E.~W.}\ \bibnamefont {Bauer}},\ }\href {\doibase
  10.1103/PhysRevB.87.144411} {\bibfield  {journal} {\bibinfo  {journal} {Phys.
  Rev. B}\ }\textbf {\bibinfo {volume} {87}},\ \bibinfo {pages} {144411}
  (\bibinfo {year} {2013})}\BibitemShut {NoStop}%
\bibitem [{\citenamefont {Ok}\ \emph {et~al.}(2017)\citenamefont {Ok},
  \citenamefont {Chen}, \citenamefont {Sigrist},\ and\ \citenamefont
  {Manske}}]{Ok17}%
  \BibitemOpen
  \bibfield  {author} {\bibinfo {author} {\bibfnamefont {S.}~\bibnamefont
  {Ok}}, \bibinfo {author} {\bibfnamefont {W.}~\bibnamefont {Chen}}, \bibinfo
  {author} {\bibfnamefont {M.}~\bibnamefont {Sigrist}}, \ and\ \bibinfo
  {author} {\bibfnamefont {D.}~\bibnamefont {Manske}},\ }\href
  {http://stacks.iop.org/0953-8984/29/i=7/a=075802} {\bibfield  {journal}
  {\bibinfo  {journal} {J. Phys. Condens. Matter}\ }\textbf {\bibinfo {volume}
  {29}},\ \bibinfo {pages} {075802} (\bibinfo {year} {2017})}\BibitemShut
  {NoStop}%
\bibitem [{\citenamefont {Moras}\ \emph {et~al.}(2015)\citenamefont {Moras},
  \citenamefont {Bihlmayer}, \citenamefont {Sheverdyaeva}, \citenamefont
  {Mahatha}, \citenamefont {Papagno}, \citenamefont {S\'anchez-Barriga},
  \citenamefont {Rader}, \citenamefont {Novinec}, \citenamefont {Gardonio},\
  and\ \citenamefont {Carbone}}]{Moras15}%
  \BibitemOpen
  \bibfield  {author} {\bibinfo {author} {\bibfnamefont {P.}~\bibnamefont
  {Moras}}, \bibinfo {author} {\bibfnamefont {G.}~\bibnamefont {Bihlmayer}},
  \bibinfo {author} {\bibfnamefont {P.~M.}\ \bibnamefont {Sheverdyaeva}},
  \bibinfo {author} {\bibfnamefont {S.~K.}\ \bibnamefont {Mahatha}}, \bibinfo
  {author} {\bibfnamefont {M.}~\bibnamefont {Papagno}}, \bibinfo {author}
  {\bibfnamefont {J.}~\bibnamefont {S\'anchez-Barriga}}, \bibinfo {author}
  {\bibfnamefont {O.}~\bibnamefont {Rader}}, \bibinfo {author} {\bibfnamefont
  {L.}~\bibnamefont {Novinec}}, \bibinfo {author} {\bibfnamefont
  {S.}~\bibnamefont {Gardonio}}, \ and\ \bibinfo {author} {\bibfnamefont
  {C.}~\bibnamefont {Carbone}},\ }\href {\doibase 10.1103/PhysRevB.91.195410}
  {\bibfield  {journal} {\bibinfo  {journal} {Phys. Rev. B}\ }\textbf {\bibinfo
  {volume} {91}},\ \bibinfo {pages} {195410} (\bibinfo {year}
  {2015})}\BibitemShut {NoStop}%
\bibitem [{\citenamefont {Carbone}\ \emph {et~al.}(2016)\citenamefont
  {Carbone}, \citenamefont {Moras}, \citenamefont {Sheverdyaeva}, \citenamefont
  {Pacil\'e}, \citenamefont {Papagno}, \citenamefont {Ferrari}, \citenamefont
  {Topwal}, \citenamefont {Vescovo}, \citenamefont {Bihlmayer}, \citenamefont
  {Freimuth}, \citenamefont {Mokrousov},\ and\ \citenamefont
  {Bl\"ugel}}]{Carbone16}%
  \BibitemOpen
  \bibfield  {author} {\bibinfo {author} {\bibfnamefont {C.}~\bibnamefont
  {Carbone}}, \bibinfo {author} {\bibfnamefont {P.}~\bibnamefont {Moras}},
  \bibinfo {author} {\bibfnamefont {P.~M.}\ \bibnamefont {Sheverdyaeva}},
  \bibinfo {author} {\bibfnamefont {D.}~\bibnamefont {Pacil\'e}}, \bibinfo
  {author} {\bibfnamefont {M.}~\bibnamefont {Papagno}}, \bibinfo {author}
  {\bibfnamefont {L.}~\bibnamefont {Ferrari}}, \bibinfo {author} {\bibfnamefont
  {D.}~\bibnamefont {Topwal}}, \bibinfo {author} {\bibfnamefont
  {E.}~\bibnamefont {Vescovo}}, \bibinfo {author} {\bibfnamefont
  {G.}~\bibnamefont {Bihlmayer}}, \bibinfo {author} {\bibfnamefont
  {F.}~\bibnamefont {Freimuth}}, \bibinfo {author} {\bibfnamefont
  {Y.}~\bibnamefont {Mokrousov}}, \ and\ \bibinfo {author} {\bibfnamefont
  {S.}~\bibnamefont {Bl\"ugel}},\ }\href {\doibase 10.1103/PhysRevB.93.125409}
  {\bibfield  {journal} {\bibinfo  {journal} {Phys. Rev. B}\ }\textbf {\bibinfo
  {volume} {93}},\ \bibinfo {pages} {125409} (\bibinfo {year}
  {2016})}\BibitemShut {NoStop}%
\bibitem [{\citenamefont {Pugh}\ and\ \citenamefont {Rostoker}(1953)}]{Pugh53}%
  \BibitemOpen
  \bibfield  {author} {\bibinfo {author} {\bibfnamefont {E.~M.}\ \bibnamefont
  {Pugh}}\ and\ \bibinfo {author} {\bibfnamefont {N.}~\bibnamefont
  {Rostoker}},\ }\href {\doibase 10.1103/RevModPhys.25.151} {\bibfield
  {journal} {\bibinfo  {journal} {Rev. Mod. Phys.}\ }\textbf {\bibinfo {volume}
  {25}},\ \bibinfo {pages} {151} (\bibinfo {year} {1953})}\BibitemShut
  {NoStop}%
\bibitem [{\citenamefont {Sinitsyn}(2008)}]{Sinitsyn08}%
  \BibitemOpen
  \bibfield  {author} {\bibinfo {author} {\bibfnamefont {N.~A.}\ \bibnamefont
  {Sinitsyn}},\ }\href {http://stacks.iop.org/0953-8984/20/i=2/a=023201}
  {\bibfield  {journal} {\bibinfo  {journal} {J. Phys. Condens. Matter}\
  }\textbf {\bibinfo {volume} {20}},\ \bibinfo {pages} {023201} (\bibinfo
  {year} {2008})}\BibitemShut {NoStop}%
\bibitem [{\citenamefont {Nagaosa}\ \emph {et~al.}(2010)\citenamefont
  {Nagaosa}, \citenamefont {Sinova}, \citenamefont {Onoda}, \citenamefont
  {MacDonald},\ and\ \citenamefont {Ong}}]{Nagaosa10}%
  \BibitemOpen
  \bibfield  {author} {\bibinfo {author} {\bibfnamefont {N.}~\bibnamefont
  {Nagaosa}}, \bibinfo {author} {\bibfnamefont {J.}~\bibnamefont {Sinova}},
  \bibinfo {author} {\bibfnamefont {S.}~\bibnamefont {Onoda}}, \bibinfo
  {author} {\bibfnamefont {A.~H.}\ \bibnamefont {MacDonald}}, \ and\ \bibinfo
  {author} {\bibfnamefont {N.~P.}\ \bibnamefont {Ong}},\ }\href {\doibase
  10.1103/RevModPhys.82.1539} {\bibfield  {journal} {\bibinfo  {journal} {Rev.
  Mod. Phys.}\ }\textbf {\bibinfo {volume} {82}},\ \bibinfo {pages} {1539}
  (\bibinfo {year} {2010})}\BibitemShut {NoStop}%
\bibitem [{\citenamefont {Kurebayashi}\ \emph {et~al.}(2014)\citenamefont
  {Kurebayashi}, \citenamefont {Sinova}, \citenamefont {Fang}, \citenamefont
  {Irvine}, \citenamefont {Skinner}, \citenamefont {Wunderlich}, \citenamefont
  {Novak}, \citenamefont {Campion}, \citenamefont {Gallagher}, \citenamefont
  {Vehstedt}, \citenamefont {Zarbo}, \citenamefont {Vyborny}, \citenamefont
  {Ferguson},\ and\ \citenamefont {Jungwirth}}]{Kurebayashi14}%
  \BibitemOpen
  \bibfield  {author} {\bibinfo {author} {\bibfnamefont {H.}~\bibnamefont
  {Kurebayashi}}, \bibinfo {author} {\bibfnamefont {J.}~\bibnamefont {Sinova}},
  \bibinfo {author} {\bibfnamefont {D.}~\bibnamefont {Fang}}, \bibinfo {author}
  {\bibfnamefont {A.~C.}\ \bibnamefont {Irvine}}, \bibinfo {author}
  {\bibfnamefont {T.~D.}\ \bibnamefont {Skinner}}, \bibinfo {author}
  {\bibfnamefont {J.}~\bibnamefont {Wunderlich}}, \bibinfo {author}
  {\bibfnamefont {V.}~\bibnamefont {Novak}}, \bibinfo {author} {\bibfnamefont
  {R.~P.}\ \bibnamefont {Campion}}, \bibinfo {author} {\bibfnamefont {B.~L.}\
  \bibnamefont {Gallagher}}, \bibinfo {author} {\bibfnamefont {E.~K.}\
  \bibnamefont {Vehstedt}}, \bibinfo {author} {\bibfnamefont {L.~P.}\
  \bibnamefont {Zarbo}}, \bibinfo {author} {\bibfnamefont {K.}~\bibnamefont
  {Vyborny}}, \bibinfo {author} {\bibfnamefont {A.~J.}\ \bibnamefont
  {Ferguson}}, \ and\ \bibinfo {author} {\bibfnamefont {T.}~\bibnamefont
  {Jungwirth}},\ }\href {\doibase 10.1038/nnano.2014.15} {\bibfield  {journal}
  {\bibinfo  {journal} {Nat. Nano.}\ }\textbf {\bibinfo {volume} {9}},\
  \bibinfo {pages} {211} (\bibinfo {year} {2014})}\BibitemShut {NoStop}%
\bibitem [{\citenamefont {Kim}\ \emph {et~al.}(2012)\citenamefont {Kim},
  \citenamefont {Seo}, \citenamefont {Ryu}, \citenamefont {Lee},\ and\
  \citenamefont {Lee}}]{Kim12}%
  \BibitemOpen
  \bibfield  {author} {\bibinfo {author} {\bibfnamefont {K.-W.}\ \bibnamefont
  {Kim}}, \bibinfo {author} {\bibfnamefont {S.-M.}\ \bibnamefont {Seo}},
  \bibinfo {author} {\bibfnamefont {J.}~\bibnamefont {Ryu}}, \bibinfo {author}
  {\bibfnamefont {K.-J.}\ \bibnamefont {Lee}}, \ and\ \bibinfo {author}
  {\bibfnamefont {H.-W.}\ \bibnamefont {Lee}},\ }\href {\doibase
  10.1103/PhysRevB.85.180404} {\bibfield  {journal} {\bibinfo  {journal} {Phys.
  Rev. B}\ }\textbf {\bibinfo {volume} {85}},\ \bibinfo {pages} {180404}
  (\bibinfo {year} {2012})}\BibitemShut {NoStop}%
\bibitem [{\citenamefont {Wang}\ and\ \citenamefont {Manchon}(2012)}]{Wang12}%
  \BibitemOpen
  \bibfield  {author} {\bibinfo {author} {\bibfnamefont {X.}~\bibnamefont
  {Wang}}\ and\ \bibinfo {author} {\bibfnamefont {A.}~\bibnamefont {Manchon}},\
  }\href {\doibase 10.1103/PhysRevLett.108.117201} {\bibfield  {journal}
  {\bibinfo  {journal} {Phys. Rev. Lett.}\ }\textbf {\bibinfo {volume} {108}},\
  \bibinfo {pages} {117201} (\bibinfo {year} {2012})}\BibitemShut {NoStop}%
\bibitem [{\citenamefont {Pesin}\ and\ \citenamefont
  {MacDonald}(2012)}]{Pesin12}%
  \BibitemOpen
  \bibfield  {author} {\bibinfo {author} {\bibfnamefont {D.~A.}\ \bibnamefont
  {Pesin}}\ and\ \bibinfo {author} {\bibfnamefont {A.~H.}\ \bibnamefont
  {MacDonald}},\ }\href {\doibase 10.1103/PhysRevB.86.014416} {\bibfield
  {journal} {\bibinfo  {journal} {Phys. Rev. B}\ }\textbf {\bibinfo {volume}
  {86}},\ \bibinfo {pages} {014416} (\bibinfo {year} {2012})}\BibitemShut
  {NoStop}%
\bibitem [{\citenamefont {Haney}\ \emph
  {et~al.}(2013{\natexlab{a}})\citenamefont {Haney}, \citenamefont {Lee},
  \citenamefont {Lee}, \citenamefont {Manchon},\ and\ \citenamefont
  {Stiles}}]{Haney13_2}%
  \BibitemOpen
  \bibfield  {author} {\bibinfo {author} {\bibfnamefont {P.~M.}\ \bibnamefont
  {Haney}}, \bibinfo {author} {\bibfnamefont {H.-W.}\ \bibnamefont {Lee}},
  \bibinfo {author} {\bibfnamefont {K.-J.}\ \bibnamefont {Lee}}, \bibinfo
  {author} {\bibfnamefont {A.}~\bibnamefont {Manchon}}, \ and\ \bibinfo
  {author} {\bibfnamefont {M.~D.}\ \bibnamefont {Stiles}},\ }\href {\doibase
  10.1103/PhysRevB.87.174411} {\bibfield  {journal} {\bibinfo  {journal} {Phys.
  Rev. B}\ }\textbf {\bibinfo {volume} {87}},\ \bibinfo {pages} {174411}
  (\bibinfo {year} {2013}{\natexlab{a}})}\BibitemShut {NoStop}%
\bibitem [{\citenamefont {Haney}\ \emph
  {et~al.}(2013{\natexlab{b}})\citenamefont {Haney}, \citenamefont {Lee},
  \citenamefont {Lee}, \citenamefont {Manchon},\ and\ \citenamefont
  {Stiles}}]{Haney13}%
  \BibitemOpen
  \bibfield  {author} {\bibinfo {author} {\bibfnamefont {P.~M.}\ \bibnamefont
  {Haney}}, \bibinfo {author} {\bibfnamefont {H.-W.}\ \bibnamefont {Lee}},
  \bibinfo {author} {\bibfnamefont {K.-J.}\ \bibnamefont {Lee}}, \bibinfo
  {author} {\bibfnamefont {A.}~\bibnamefont {Manchon}}, \ and\ \bibinfo
  {author} {\bibfnamefont {M.~D.}\ \bibnamefont {Stiles}},\ }\href {\doibase
  10.1103/PhysRevB.88.214417} {\bibfield  {journal} {\bibinfo  {journal} {Phys.
  Rev. B}\ }\textbf {\bibinfo {volume} {88}},\ \bibinfo {pages} {214417}
  (\bibinfo {year} {2013}{\natexlab{b}})}\BibitemShut {NoStop}%
\bibitem [{\citenamefont {Qiu}\ \emph {et~al.}(2014)\citenamefont {Qiu},
  \citenamefont {Deorani}, \citenamefont {Narayanapillai}, \citenamefont {Lee},
  \citenamefont {Lee}, \citenamefont {Lee},\ and\ \citenamefont
  {Yang}}]{Qiu14}%
  \BibitemOpen
  \bibfield  {author} {\bibinfo {author} {\bibfnamefont {X.}~\bibnamefont
  {Qiu}}, \bibinfo {author} {\bibfnamefont {P.}~\bibnamefont {Deorani}},
  \bibinfo {author} {\bibfnamefont {K.}~\bibnamefont {Narayanapillai}},
  \bibinfo {author} {\bibfnamefont {K.-S.}\ \bibnamefont {Lee}}, \bibinfo
  {author} {\bibfnamefont {K.-J.}\ \bibnamefont {Lee}}, \bibinfo {author}
  {\bibfnamefont {H.-W.}\ \bibnamefont {Lee}}, \ and\ \bibinfo {author}
  {\bibfnamefont {H.}~\bibnamefont {Yang}},\ }\href {\doibase
  10.1038/srep04491} {\bibfield  {journal} {\bibinfo  {journal} {Sci. Rep.}\
  }\textbf {\bibinfo {volume} {4}},\ \bibinfo {pages} {4491} (\bibinfo {year}
  {2014})}\BibitemShut {NoStop}%
\bibitem [{\citenamefont {Pauyac}\ \emph {et~al.}(2013)\citenamefont {Pauyac},
  \citenamefont {Wang}, \citenamefont {Chshiev},\ and\ \citenamefont
  {Manchon}}]{Pauyac13}%
  \BibitemOpen
  \bibfield  {author} {\bibinfo {author} {\bibfnamefont {C.~O.}\ \bibnamefont
  {Pauyac}}, \bibinfo {author} {\bibfnamefont {X.}~\bibnamefont {Wang}},
  \bibinfo {author} {\bibfnamefont {M.}~\bibnamefont {Chshiev}}, \ and\
  \bibinfo {author} {\bibfnamefont {A.}~\bibnamefont {Manchon}},\ }\href
  {\doibase 10.1063/1.4812663} {\bibfield  {journal} {\bibinfo  {journal}
  {Applied Physics Letters}\ }\textbf {\bibinfo {volume} {102}},\ \bibinfo
  {pages} {252403} (\bibinfo {year} {2013})}\BibitemShut {NoStop}%
\bibitem [{\citenamefont {Lee}\ \emph {et~al.}(2015)\citenamefont {Lee},
  \citenamefont {Go}, \citenamefont {Manchon}, \citenamefont {Haney},
  \citenamefont {Stiles}, \citenamefont {Lee},\ and\ \citenamefont
  {Lee}}]{Lee15}%
  \BibitemOpen
  \bibfield  {author} {\bibinfo {author} {\bibfnamefont {K.-S.}\ \bibnamefont
  {Lee}}, \bibinfo {author} {\bibfnamefont {D.}~\bibnamefont {Go}}, \bibinfo
  {author} {\bibfnamefont {A.}~\bibnamefont {Manchon}}, \bibinfo {author}
  {\bibfnamefont {P.~M.}\ \bibnamefont {Haney}}, \bibinfo {author}
  {\bibfnamefont {M.~D.}\ \bibnamefont {Stiles}}, \bibinfo {author}
  {\bibfnamefont {H.-W.}\ \bibnamefont {Lee}}, \ and\ \bibinfo {author}
  {\bibfnamefont {K.-J.}\ \bibnamefont {Lee}},\ }\href {\doibase
  10.1103/PhysRevB.91.144401} {\bibfield  {journal} {\bibinfo  {journal} {Phys.
  Rev. B}\ }\textbf {\bibinfo {volume} {91}},\ \bibinfo {pages} {144401}
  (\bibinfo {year} {2015})}\BibitemShut {NoStop}%
\bibitem [{\citenamefont {Bass}\ and\ \citenamefont {Pratt~Jr}(2007)}]{Bass07}%
  \BibitemOpen
  \bibfield  {author} {\bibinfo {author} {\bibfnamefont {J.}~\bibnamefont
  {Bass}}\ and\ \bibinfo {author} {\bibfnamefont {W.~P.}\ \bibnamefont
  {Pratt~Jr}},\ }\href {http://stacks.iop.org/0953-8984/19/i=18/a=183201}
  {\bibfield  {journal} {\bibinfo  {journal} {J. Phys. Condens. Matter}\
  }\textbf {\bibinfo {volume} {19}},\ \bibinfo {pages} {183201} (\bibinfo
  {year} {2007})}\BibitemShut {NoStop}%
\bibitem [{\citenamefont {Chen}\ \emph {et~al.}(2015)\citenamefont {Chen},
  \citenamefont {Sigrist}, \citenamefont {Sinova},\ and\ \citenamefont
  {Manske}}]{Chen15_STT}%
  \BibitemOpen
  \bibfield  {author} {\bibinfo {author} {\bibfnamefont {W.}~\bibnamefont
  {Chen}}, \bibinfo {author} {\bibfnamefont {M.}~\bibnamefont {Sigrist}},
  \bibinfo {author} {\bibfnamefont {J.}~\bibnamefont {Sinova}}, \ and\ \bibinfo
  {author} {\bibfnamefont {D.}~\bibnamefont {Manske}},\ }\href {\doibase
  10.1103/PhysRevLett.115.217203} {\bibfield  {journal} {\bibinfo  {journal}
  {Phys. Rev. Lett.}\ }\textbf {\bibinfo {volume} {115}},\ \bibinfo {pages}
  {217203} (\bibinfo {year} {2015})}\BibitemShut {NoStop}%
\bibitem [{\citenamefont {Chen}\ \emph {et~al.}(2016)\citenamefont {Chen},
  \citenamefont {Sigrist},\ and\ \citenamefont
  {Manske}}]{Chen16_quantum_tunneling}%
  \BibitemOpen
  \bibfield  {author} {\bibinfo {author} {\bibfnamefont {W.}~\bibnamefont
  {Chen}}, \bibinfo {author} {\bibfnamefont {M.}~\bibnamefont {Sigrist}}, \
  and\ \bibinfo {author} {\bibfnamefont {D.}~\bibnamefont {Manske}},\ }\href
  {\doibase 10.1103/PhysRevB.94.104412} {\bibfield  {journal} {\bibinfo
  {journal} {Phys. Rev. B}\ }\textbf {\bibinfo {volume} {94}},\ \bibinfo
  {pages} {104412} (\bibinfo {year} {2016})}\BibitemShut {NoStop}%
\bibitem [{\citenamefont {Kim}\ \emph {et~al.}(2013)\citenamefont {Kim},
  \citenamefont {Sinha}, \citenamefont {Hayashi}, \citenamefont {Yamanouchi},
  \citenamefont {Fukami}, \citenamefont {Suzuki}, \citenamefont {Mitani},\ and\
  \citenamefont {Ohno}}]{Kim13}%
  \BibitemOpen
  \bibfield  {author} {\bibinfo {author} {\bibfnamefont {J.}~\bibnamefont
  {Kim}}, \bibinfo {author} {\bibfnamefont {J.}~\bibnamefont {Sinha}}, \bibinfo
  {author} {\bibfnamefont {M.}~\bibnamefont {Hayashi}}, \bibinfo {author}
  {\bibfnamefont {M.}~\bibnamefont {Yamanouchi}}, \bibinfo {author}
  {\bibfnamefont {S.}~\bibnamefont {Fukami}}, \bibinfo {author} {\bibfnamefont
  {T.}~\bibnamefont {Suzuki}}, \bibinfo {author} {\bibfnamefont
  {S.}~\bibnamefont {Mitani}}, \ and\ \bibinfo {author} {\bibfnamefont
  {H.}~\bibnamefont {Ohno}},\ }\href {\doibase 10.1038/nmat3522} {\bibfield
  {journal} {\bibinfo  {journal} {Nat. Mater.}\ }\textbf {\bibinfo {volume}
  {12}},\ \bibinfo {pages} {240} (\bibinfo {year} {2013})}\BibitemShut
  {NoStop}%
\bibitem [{\citenamefont {Lee}\ \emph {et~al.}(2011)\citenamefont {Lee},
  \citenamefont {Kim}, \citenamefont {Ryu}, \citenamefont {Moon}, \citenamefont
  {Yun}, \citenamefont {Gim}, \citenamefont {Lee}, \citenamefont {Shin},
  \citenamefont {Lee},\ and\ \citenamefont {Choe}}]{Lee11}%
  \BibitemOpen
  \bibfield  {author} {\bibinfo {author} {\bibfnamefont {J.-C.}\ \bibnamefont
  {Lee}}, \bibinfo {author} {\bibfnamefont {K.-J.}\ \bibnamefont {Kim}},
  \bibinfo {author} {\bibfnamefont {J.}~\bibnamefont {Ryu}}, \bibinfo {author}
  {\bibfnamefont {K.-W.}\ \bibnamefont {Moon}}, \bibinfo {author}
  {\bibfnamefont {S.-J.}\ \bibnamefont {Yun}}, \bibinfo {author} {\bibfnamefont
  {G.-H.}\ \bibnamefont {Gim}}, \bibinfo {author} {\bibfnamefont {K.-S.}\
  \bibnamefont {Lee}}, \bibinfo {author} {\bibfnamefont {K.-H.}\ \bibnamefont
  {Shin}}, \bibinfo {author} {\bibfnamefont {H.-W.}\ \bibnamefont {Lee}}, \
  and\ \bibinfo {author} {\bibfnamefont {S.-B.}\ \bibnamefont {Choe}},\ }\href
  {\doibase 10.1103/PhysRevLett.107.067201} {\bibfield  {journal} {\bibinfo
  {journal} {Phys. Rev. Lett.}\ }\textbf {\bibinfo {volume} {107}},\ \bibinfo
  {pages} {067201} (\bibinfo {year} {2011})}\BibitemShut {NoStop}%
\bibitem [{\citenamefont {Miron}\ \emph
  {et~al.}(2011{\natexlab{b}})\citenamefont {Miron}, \citenamefont {Moore},
  \citenamefont {Szambolics}, \citenamefont {Buda-Prejbeanu}, \citenamefont
  {Auffret}, \citenamefont {Rodmacq}, \citenamefont {Pizzini}, \citenamefont
  {Vogel}, \citenamefont {Bonfim}, \citenamefont {Schuhl},\ and\ \citenamefont
  {Gaudin}}]{Miron11_2}%
  \BibitemOpen
  \bibfield  {author} {\bibinfo {author} {\bibfnamefont {I.~M.}\ \bibnamefont
  {Miron}}, \bibinfo {author} {\bibfnamefont {T.}~\bibnamefont {Moore}},
  \bibinfo {author} {\bibfnamefont {H.}~\bibnamefont {Szambolics}}, \bibinfo
  {author} {\bibfnamefont {L.~D.}\ \bibnamefont {Buda-Prejbeanu}}, \bibinfo
  {author} {\bibfnamefont {S.}~\bibnamefont {Auffret}}, \bibinfo {author}
  {\bibfnamefont {B.}~\bibnamefont {Rodmacq}}, \bibinfo {author} {\bibfnamefont
  {S.}~\bibnamefont {Pizzini}}, \bibinfo {author} {\bibfnamefont
  {J.}~\bibnamefont {Vogel}}, \bibinfo {author} {\bibfnamefont
  {M.}~\bibnamefont {Bonfim}}, \bibinfo {author} {\bibfnamefont
  {A.}~\bibnamefont {Schuhl}}, \ and\ \bibinfo {author} {\bibfnamefont
  {G.}~\bibnamefont {Gaudin}},\ }\href {\doibase 10.1038/nmat3020} {\bibfield
  {journal} {\bibinfo  {journal} {Nat. Mater.}\ }\textbf {\bibinfo {volume}
  {10}},\ \bibinfo {pages} {419} (\bibinfo {year}
  {2011}{\natexlab{b}})}\BibitemShut {NoStop}%
\bibitem [{\citenamefont {Emori}\ \emph {et~al.}(2013)\citenamefont {Emori},
  \citenamefont {Bauer}, \citenamefont {Ahn}, \citenamefont {Martinez},\ and\
  \citenamefont {Beach}}]{Emori13}%
  \BibitemOpen
  \bibfield  {author} {\bibinfo {author} {\bibfnamefont {S.}~\bibnamefont
  {Emori}}, \bibinfo {author} {\bibfnamefont {U.}~\bibnamefont {Bauer}},
  \bibinfo {author} {\bibfnamefont {S.-M.}\ \bibnamefont {Ahn}}, \bibinfo
  {author} {\bibfnamefont {E.}~\bibnamefont {Martinez}}, \ and\ \bibinfo
  {author} {\bibfnamefont {G.~S.~D.}\ \bibnamefont {Beach}},\ }\href {\doibase
  10.1038/nmat3675} {\bibfield  {journal} {\bibinfo  {journal} {Nat. Mater.}\
  }\textbf {\bibinfo {volume} {12}},\ \bibinfo {pages} {611} (\bibinfo {year}
  {2013})}\BibitemShut {NoStop}%
\bibitem [{\citenamefont {Kato}\ \emph {et~al.}(2004)\citenamefont {Kato},
  \citenamefont {Myers}, \citenamefont {Gossard},\ and\ \citenamefont
  {Awschalom}}]{Kato04}%
  \BibitemOpen
  \bibfield  {author} {\bibinfo {author} {\bibfnamefont {Y.~K.}\ \bibnamefont
  {Kato}}, \bibinfo {author} {\bibfnamefont {R.~C.}\ \bibnamefont {Myers}},
  \bibinfo {author} {\bibfnamefont {A.~C.}\ \bibnamefont {Gossard}}, \ and\
  \bibinfo {author} {\bibfnamefont {D.~D.}\ \bibnamefont {Awschalom}},\ }\href
  {\doibase 10.1126/science.1105514} {\bibfield  {journal} {\bibinfo  {journal}
  {Science}\ }\textbf {\bibinfo {volume} {306}},\ \bibinfo {pages} {1910}
  (\bibinfo {year} {2004})}\BibitemShut {NoStop}%
\bibitem [{\citenamefont {Lavrijsen}\ \emph {et~al.}(2012)\citenamefont
  {Lavrijsen}, \citenamefont {Haazen}, \citenamefont {Murè}, \citenamefont
  {Franken}, \citenamefont {Kohlhepp}, \citenamefont {Swagten},\ and\
  \citenamefont {Koopmans}}]{Lavrijsen12}%
  \BibitemOpen
  \bibfield  {author} {\bibinfo {author} {\bibfnamefont {R.}~\bibnamefont
  {Lavrijsen}}, \bibinfo {author} {\bibfnamefont {P.~P.~J.}\ \bibnamefont
  {Haazen}}, \bibinfo {author} {\bibfnamefont {E.}~\bibnamefont {Murè}},
  \bibinfo {author} {\bibfnamefont {J.~H.}\ \bibnamefont {Franken}}, \bibinfo
  {author} {\bibfnamefont {J.~T.}\ \bibnamefont {Kohlhepp}}, \bibinfo {author}
  {\bibfnamefont {H.~J.~M.}\ \bibnamefont {Swagten}}, \ and\ \bibinfo {author}
  {\bibfnamefont {B.}~\bibnamefont {Koopmans}},\ }\href {\doibase
  10.1063/1.4732083} {\bibfield  {journal} {\bibinfo  {journal} {Applied
  Physics Letters}\ }\textbf {\bibinfo {volume} {100}},\ \bibinfo {pages}
  {262408} (\bibinfo {year} {2012})}\BibitemShut {NoStop}%
\bibitem [{\citenamefont {Wang}\ \emph {et~al.}(2015)\citenamefont {Wang},
  \citenamefont {Chen},\ and\ \citenamefont {Zhang}}]{Wang15}%
  \BibitemOpen
  \bibfield  {author} {\bibinfo {author} {\bibfnamefont {Y.}~\bibnamefont
  {Wang}}, \bibinfo {author} {\bibfnamefont {W.-Q.}\ \bibnamefont {Chen}}, \
  and\ \bibinfo {author} {\bibfnamefont {F.-C.}\ \bibnamefont {Zhang}},\ }\href
  {http://stacks.iop.org/1367-2630/17/i=5/a=053012} {\bibfield  {journal}
  {\bibinfo  {journal} {New J. Phys.}\ }\textbf {\bibinfo {volume} {17}},\
  \bibinfo {pages} {053012} (\bibinfo {year} {2015})}\BibitemShut {NoStop}%
\end{thebibliography}%

\end{document}